\documentclass[conference]{IEEEtran}
\IEEEoverridecommandlockouts
\usepackage{cite}
\usepackage{amsmath,amssymb,amsfonts}
\usepackage{algorithmic}
\usepackage{graphicx}
\usepackage{textcomp}
\usepackage{xcolor}
\usepackage{url}
\usepackage{float}
\usepackage{subfigure}
\usepackage[ruled,vlined]{algorithm2e}
\newcommand{\omt}[1]{}
\newtheorem{theorem}{Theorem}
\newcommand{\proof}[1]{{\em Proof.} {#1} \rule{2mm}{2mm}}

\def\BibTeX{{\rm B\kern-.05em{\sc i\kern-.025em b}\kern-.08em
    T\kern-.1667em\lower.7ex\hbox{E}\kern-.125emX}}
\begin{document}

\title{Hypergraph Ego-networks and Their Temporal Evolution

}

\author{\IEEEauthorblockN{
Cazamere Comrie, Jon Kleinberg}
\IEEEauthorblockA{\textit{Cornell University} \\
\{clc348, kleinberg\}@cornell.edu}
}

\maketitle

\begin{abstract}
Interactions involving multiple objects simultaneously are ubiquitous across many domains. The systems these interactions inhabit can be modelled using hypergraphs, a generalization of traditional  graphs in which each edge can connect any number of nodes. Analyzing the global and static properties of these hypergraphs has led to a plethora of novel findings regarding how these modelled systems are structured. However, less is known about the localized structure of these systems and how they evolve over time. \par
In this paper, we propose the study of hypergraph ego-networks, a structure that can be used to model higher-order interactions involving a single node. We also propose the temporal reconstruction of hypergraph ego-networks as a benchmark problem for models that aim to predict the local temporal structure of hypergraphs. By combining a deep learning binary classifier with a hill-climbing algorithm, we will present a model for reconstructing hypergraph ego-networks by incorporating structural patterns found across multiple domains.
\end{abstract}

\maketitle

\section{Introduction}
\noindent Interactions involving multiple objects simultaneously are ubiquitous across many domains: academic papers tend to have multiple co-authors, emails are sent to multiple recipients, and bills in congress are co-sponsored by multiple members. A growing body of research has been dedicated to understanding the structure of these higher-order interactions \cite{benson2018simplicial, benson2016higher, benson2018sequences, do2020structural, kook2020evolution}, seeking to explore these interactions by analyzing the macroscopic trends of the interactions within these systems. \par
These systems can be modelled using hypergraphs \cite{berge1984hypergraphs}. As generalizations of graphs, hypergraphs are composed of nodes and hyperedges, where each edge can contain any number of nodes. For example, a hypergraph modelling a co-authorship network would represent each author as a node, and each hyperedge would represent a paper co-authored by a set of authors. Whereas a typical graph could only model the pairwise interactions among this set of authors, a hypergraph allows for authors to interact with multiple other authors simultaneously, and thus captures these higher-order interactions  effectively. \par
There are two dichotomies worth considering when discussing the structure of hypergraphs. The first is whether a hypergraph is static or temporal. Static hypergraphs have been studied extensively, where researchers have analyzed the structure of the hypergraph at a specific moment in time. However, very little is known about temporal hypergraphs, which can be viewed as an ordered, timestamped sequence of hypergraphs. Kook et al. \cite{kook2020evolution} have recently considered aggregate patterns found in real-world temporal hypergraphs. \par
The second dichotomy is whether a hypergraph property is global or local. Analyzing the global properties of  hypergraphs has lead to a plethora of novel findings regarding how these modelled systems are structured. For example, \cite{do2020structural} observes that large real-world hypergraphs at the macroscopic level have similar well-known properties to real-world dyadic graphs, such as a giant connected component and a heavy-tailed degree distribution. However, little is known about the localized structure of higher-order systems. \par
In this work we aim to fill these gaps in knowledge by studying the local, temporal structure of hypergraphs.  We proceed by drawing an analogy to corresponding structures in traditional (pairwise) graphs, where work has been done on analyzing all the interactions involving a single node \cite{aiello2017evolution, arnaboldi2017online, arnaboldi2016analysis}. These interactions centering on one node are commonly modelled using \emph{ego-networks}, the network of pairwise interactions among the neighbors of a single node. Ego-networks in dyadic graphs are used to understand not only the structure of local interactions, but how these local interactions influence the behavior of the global system they inhabit. \par
What would be the analogue of an ego-network in the hypergraph context?  To better contextualize this, as an example consider the complete history of all papers written by a single author, where each paper is represented as a hyperedge containing all authors of each paper. This in turn forms the author's ego-network. It is then interesting to ask whether there are any recurring patterns to the order in which hyperedges appear in the ego-network, and if there are fundamental properties of temporal higher-order interactions that lead to these patterns. By better understanding the temporal nature of these group interactions, we can gain new insights on how ego-networks grow across different domains, and how these local patterns inform global properties of hypergraphs. \par
Yet, simply proposing models for understanding the temporal evolution of localized hypergraphs is not enough. Without some sort of benchmark problem, it is very difficult to know which models are better than others. Any effective model that attempts to understand the temporal evolution of these aforementioned systems should be inherently predictive. Therefore, there is significant value in the creation of a benchmark problem where one can evaluate their evolutionary models. \par
In this paper, we propose the study of hypergraph ego-networks, a structure that can be analogously used to model higher-order interactions involving a single node and its neighbors. We propose the temporal reconstruction of hypergraph ego-networks as a benchmark problem for models that aim to predict the local temporal structure of hypergraphs. We can convey the problem as follows. Suppose we are given a node $v$ that is an element in $m$ hyperedges, and a list of hyperedges $e_1, e_2, ..., e_m$ not sorted by time. How accurately can we predict the order in which the hyperedges arrived? Is there an algorithm that can significantly beat simple baselines derived from random ordering? Here we present a model that outperforms this baseline by incorporating structural patterns found in hypergraph ego-networks across multiple domains. \par
We demonstrate the effectiveness of our model in three different datasets, capturing higher-order structure in three distinct domains. The first is coauth-DBLP, a publication dataset where nodes represent authors and hyperedges represent a paper co-authored by a set of authors. Next is email-Avocado, a collection of emails where each node represents an email account, and each hyperedge contains all nodes who received the same email together with the sender of the email. The third dataset is threads-ask-ubuntu, where each node represents a user, and each hyperedge contains all users answering a question on a forum. These datasets are particularly useful to study for this problem as they contain entire lifetimes of user activity. The entire history of every interaction for each user is recorded in each dataset, allowing us to accurately see when nodes are first added to ego-networks, and to analyze the temporal structure of each user's ego-network. \par

\begin{figure*}
\subfigure[Simplex Dataset]{\includegraphics[width=4cm]{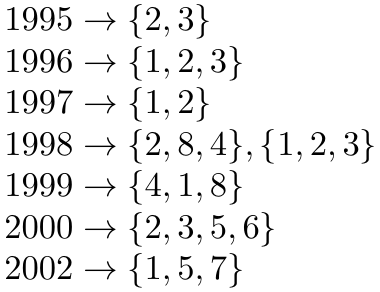}}
\subfigure[Star]{\includegraphics[width=4.6cm]{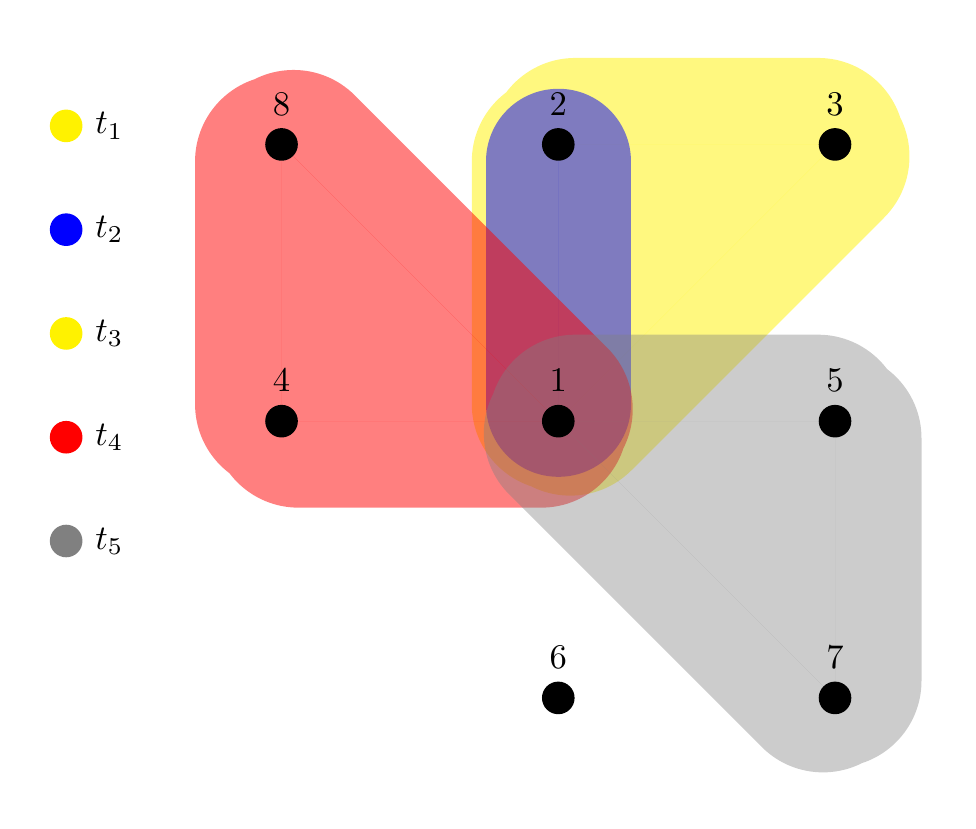}}
\subfigure[Radial]{\includegraphics[width=4.6cm]{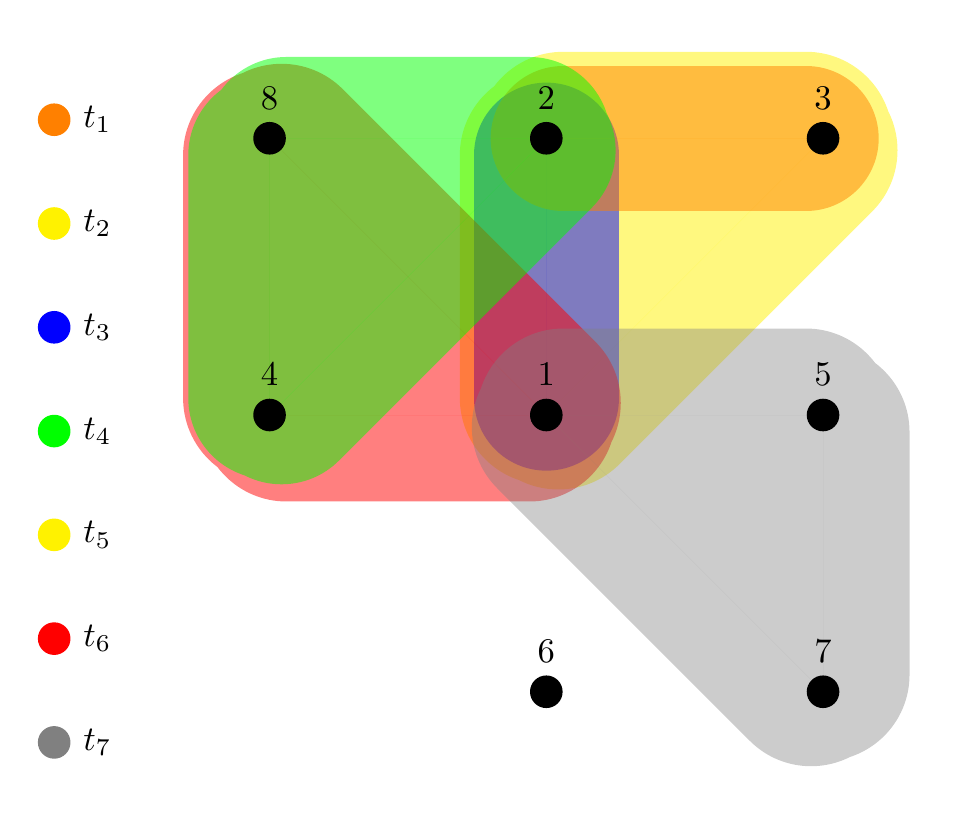}}
\subfigure[Contracted]{\includegraphics[width=4.3cm]{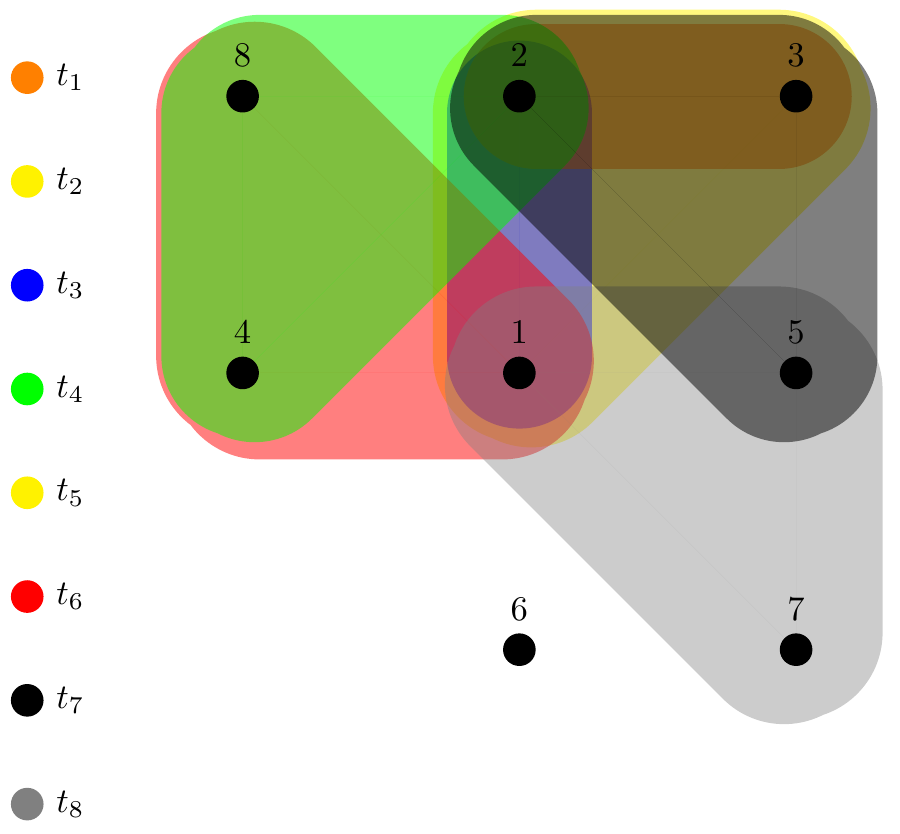}}
\caption{\textbf{Example of three different hypergraph ego-networks of user node $1$ in the domain of co-authorship. (a) Higher-order network consisting of eight timestamp papers modelled as simplices on eight authors (nodes). Note that two papers were published in the year 1998, which we will add one by one in the horizontal order they appear in the dataset. (b) Star ego-network of user node $1$. The star ego-network consists of all simplices in our dataset that include the user node. Therefore, in ordinal time, the star ego-network of $1$ will be: $t_1 \rightarrow \{1, 2, 3\}$,  $t_2 \rightarrow \{1, 2\}$, $t_3 \rightarrow \{1, 2, 3\}$, $t_4 \rightarrow \{4, 1, 8\}$, $t_5 \rightarrow \{1, 5, 7\}$. (c) Radial ego-network of $1$. Since the radial ego-network additionally includes all simplices from the dataset where all nodes are alters, the two simplices $\{2, 3\}$ and $\{2, 8, 4\}$ are added into the ego-network. As a result, the radial ego-network is: $t_1 \rightarrow \{2, 3\}$,  $t_2 \rightarrow \{1, 2, 3\}$, $t_3 \rightarrow \{1, 2\}$, $t_4 \rightarrow \{2, 8, 4\}$, $t_5 \rightarrow \{1, 2, 3\}$, $t_6 \rightarrow \{4, 1, 8\}$, $t_7 \rightarrow \{1, 5, 7\}$. (d) Contracted ego-network of $1$. The contracted ego-network consists of the intersection of every simplex in the dataset with the set $\{A(u) \cup u\}$, where $u$ is the user node (in this case $1$) and $A(u)$ is the set of all alters of $u$. In our example, this means that we will additionally take the intersection of the simplex $\{2, 3, 5, 6\}$ and  $\{A(u) \cup u\}$. As the node $6$ is not an alter of the user node $1$, we will add the simplex $\{2, 3, 5\}$ instead, removing $6$ from the set. Therefore, the contracted ego-network is: $t_1 \rightarrow \{2, 3\}$,  $t_2 \rightarrow \{1, 2, 3\}$, $t_3 \rightarrow \{1, 2\}$, $t_4 \rightarrow \{2, 8, 4\}$, $t_5 \rightarrow \{1, 2, 3\}$, $t_6 \rightarrow \{4, 1, 8\}$, $t_7 \rightarrow \{2, 3, 5\}$, $t_8 \rightarrow \{1, 5, 7\}$.}}
\label{fig:first-example}
\end{figure*}

Before we describe our model, we will define a few basic constraints for our temporal reconstruction problem. Firstly, a hyperedge is defined to be \textit{non-trivial} if it contains at least two nodes, and is \textit{trivial} otherwise. The hypergraph ego-networks we will be analyzing will exclude all trivial hyperedges, as these do not capture information regarding a node's higher-order interactions with others. Secondly, the \textit{length} or \textit{size} of an ego-network is defined as the total number of hyperedges in the ego-network. We only analyze ego-networks of at least some minimum length (typically 20 or greater for coauth-DBLP and 10 or greater for email-Avocado and threads-ask-ubuntu). This is so that our model can observe a sufficient number of higher-order interactions before it makes any predictions. We will only analyze hypergraph ego-networks that grow through the addition of hyperedges, so no hyperedges will be removed from a growing ego-network. We leave the temporal analysis of shrinking ego-networks for future work. Finally, we will ignore all ego-networks where the majority of hyperedges are identical and where the ego-network has less than 10 neighbors (except for email-Avocado), as any model that attempts to sort these ego-networks will do well. \par
Our approach is as follows. Firstly, we propose a supervised deep learning method to learn if a given ego-network is correctly ordered in time. We do this by training on ego-networks datasets where half of the ego-networks are correctly ordered. Our method requires only a few crucial combinatorial features in order to perform significantly well, and generalizes across our datasets without any changes in learning parameters. We also find that temporal features are the most significant for prediction. On all datasets, our model significantly outperforms baselines derived from random ordering. \par
Next, we define a hill climbing algorithm that is given our learning method and a shuffled ego-network as input. Each iteration of the algorithm swaps every pair of hyperedges and applies the supervised method to each ordering. Out of all the orderings that have increased likelihood of being a correct ordering, the algorithm chooses one at random. This then becomes the input ordering for the next iteration. This process is repeated until no swaps improve likelihood, in which case the algorithm saves the current ordering, and repeats the above process on another randomized ordering. After a certain amount of further attempts, the algorithm then returns the stored ordering with highest likelihood. Experimental results show that our model significantly outperforms multiple baselines on each dataset. Thus, the model is able to accurately reconstruct hypergraph ego-networks across domains.\par

\section{Basic Definitions}
\noindent \textbf{Hypergraphs:} \textit{Hypergraphs} are generalizations of traditional pairwise graphs where an edge can connect any number of nodes, whereas an edge in a graph only connect two nodes together \cite{berge1984hypergraphs}. We refer to these edges as hyperedges. More formally, a hypergraph $G = (V, E)$ consists of a set of \textit{nodes} $V$ and a set of subsets of $V$ known as $E$ (the set of \textit{hyperedges}). Each hyperedge $e \in E$ contains a number of nodes $|e|$, which we refer to as its \textit{size}. Each node $v \in V$ can belong to multiple hyperedges, and the number of hyperedges a particular node belong to is known as the \textit{degree} of the node. \par

\noindent \textbf{Simplex:} In this paper, we will be looking at temporal hypergraphs where each hyperedge is associated with a particular timestamp. From this point on, we will be referring to these hyperedges as \textit{simplices}, and use $S$ to denote the set of all simplices. We define the \textit{size} of a simplex to be the number of nodes in the simplex. We also define a simplex to be \textit{trivial} if the simplex is of size less than 2, and is \textit{non-trivial} otherwise.\par

\noindent \textbf{Hypergraph Ego-network:} The \textit{hypergraph ego-network} $E$ of a node $u$ is the set of simplices that represent the interactions among $u$'s neighbors. We refer to $u$ as the \textit{user node} or the \textit{ego}, and we refer to $u$'s neighbors as \textit{alters}. For the sake of brevity, we will mostly refer to hypergraph ego-networks from this point on as just \textit{ego-networks}. We define the \textit{length} of an ego-network to be the number of simplices in it.  Because there are multiple natural ways to represent the interactions among a node $u$'s neighbors --- for example, whether all interactions must involve $u$ or whether some can take place only among the neighbors --- we propose definitions for three natural, distinct types of hypergraph ego-networks: \textit{star}, \textit{radial}, and \textit{contracted ego-networks}. We define them here, and give an example of the three types in Figure \ref{fig:first-example}. \par

\noindent \textbf{Star Ego-network:} If $S$ is the set of all simplices, and $u$ is the user node, the \textit{star ego-network} $T(u)$ is defined as follows:
\begin{displaymath}T(u) = \{s: (u \in s)\}, \forall s \in S\end{displaymath}
In other words, the star ego-network is composed of all simplices that include the user node. We refer to simplices of this type as \textit{user simplices}. This is the simplest type of ego-network we will be working with in this paper, as it does not model any interactions between the alter nodes except for those that involve the user node $u$. Note that the richness of this structure really manifests itself only in structures with hyperedges on at least three nodes; it is not nearly as interesting in the dyadic case, where pairwise edges only connect a user node to its neighbors. Star hypergraph ego-networks are still able to model interactions among alters, provided those interactions take place within a user simplex.
\par

\noindent \textbf{Radial Ego-network:} If $S$ is the set of all simplices, $u$ is the user node, and $A(u)$ is the set of all alters of $u$, the \textit{radial ego-network} $R(u)$ is defined as follows:
\begin{displaymath}R(u) = \{s: s \subseteq \{A(u) \cup \{u\}\}\}, \forall s \in S\end{displaymath}
The radial ego-network is composed of all simplices where every node in the simplex is either the user node or an alter of the user node. With the radial ego-network, we are able to include interactions among alters, as some of these simplices may not include the user node at all. These simplices that are composed entirely of alters we define as \textit{alter simplices}.\par
\noindent \textbf{Contracted Ego-network:}  If $S$ is the set of all simplices, $u$ is the user node, and $A(u)$ is the set of all alters, the \textit{contracted ego-network} $C(u)$ is defined as follow:

\begin{displaymath}C(u) = \{s \cap \{A(u) \cup \{u\}\} \}, \forall s \in S\end{displaymath}

The contracted ego-network is the intersection of each simplex in $S$ with the set $\{A(u) \cup \{u\}\}$. The contracted ego-network captures more alter simplices than the radial ego-network, as it includes subsets of simplices that interact with nodes outside of $u$'s ego-network. \par
The above definitions lead to a key structural property of the three ego-network types. For some user node $u$:
\begin{displaymath}T(u) \subseteq R(u) \subseteq C(u)\end{displaymath}
It is important to note the scope of the information available to an ego-network. Each ego-network only has access to the local information of each of the alters, which includes all the interactions of that alter in the ego-network. The ego-network does not have access to any of their alter's interactions that take place outside of the ego-network. In other words, when we eventually apply our supervised learning model to an ego-network, the model will not have access to information such as the structure of every alter's individual ego-network. We leave the incorporation of such information for future work. \par

\begin{figure*}[t]
\subfigure[Star - coauth-DBLP]{\includegraphics[width=6cm]{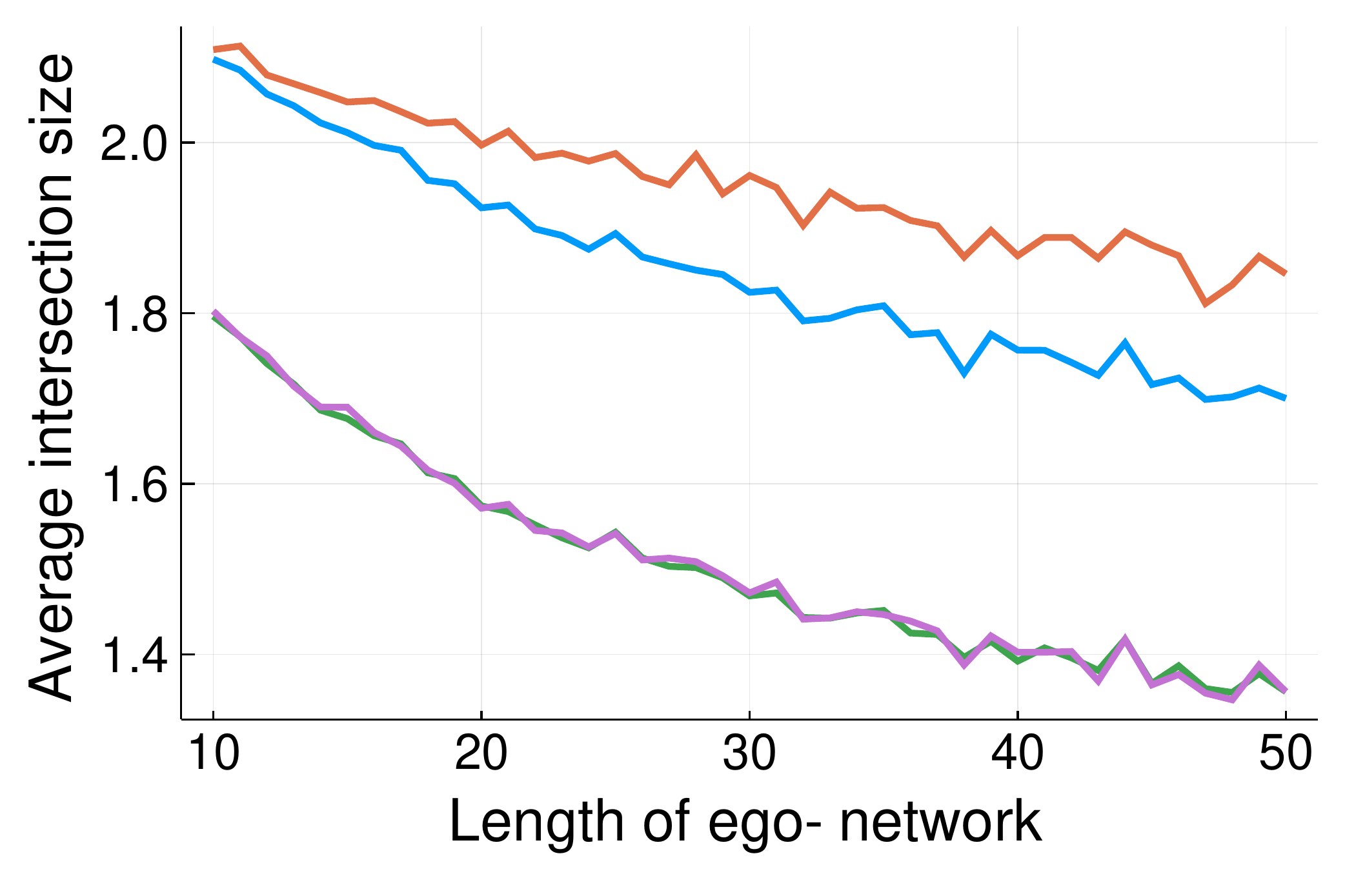}}
\subfigure[Radial - coauth-DBLP]{\includegraphics[width=6cm]{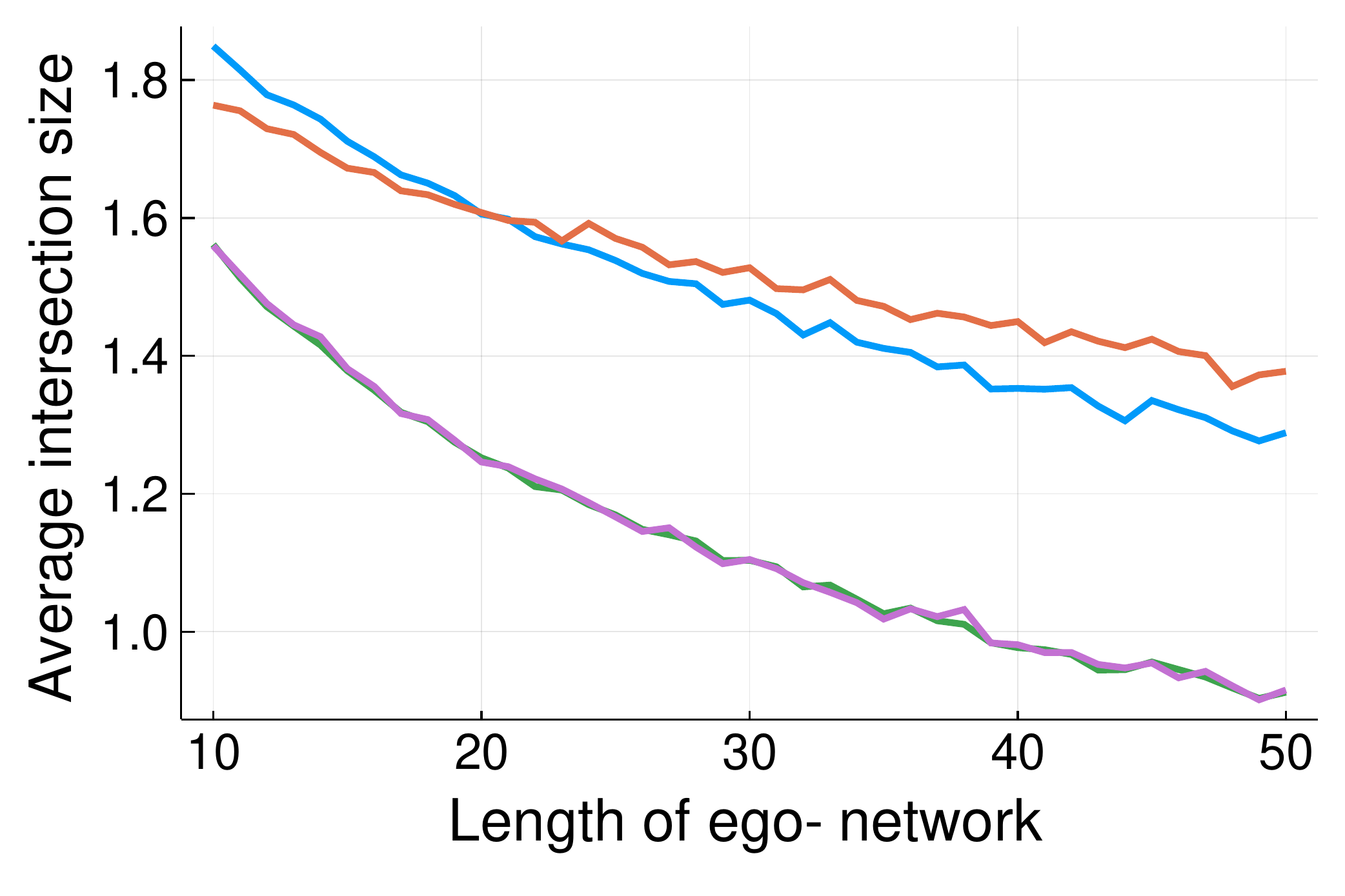}}
\subfigure[Contracted - coauth-DBLP]{\includegraphics[width=6cm]{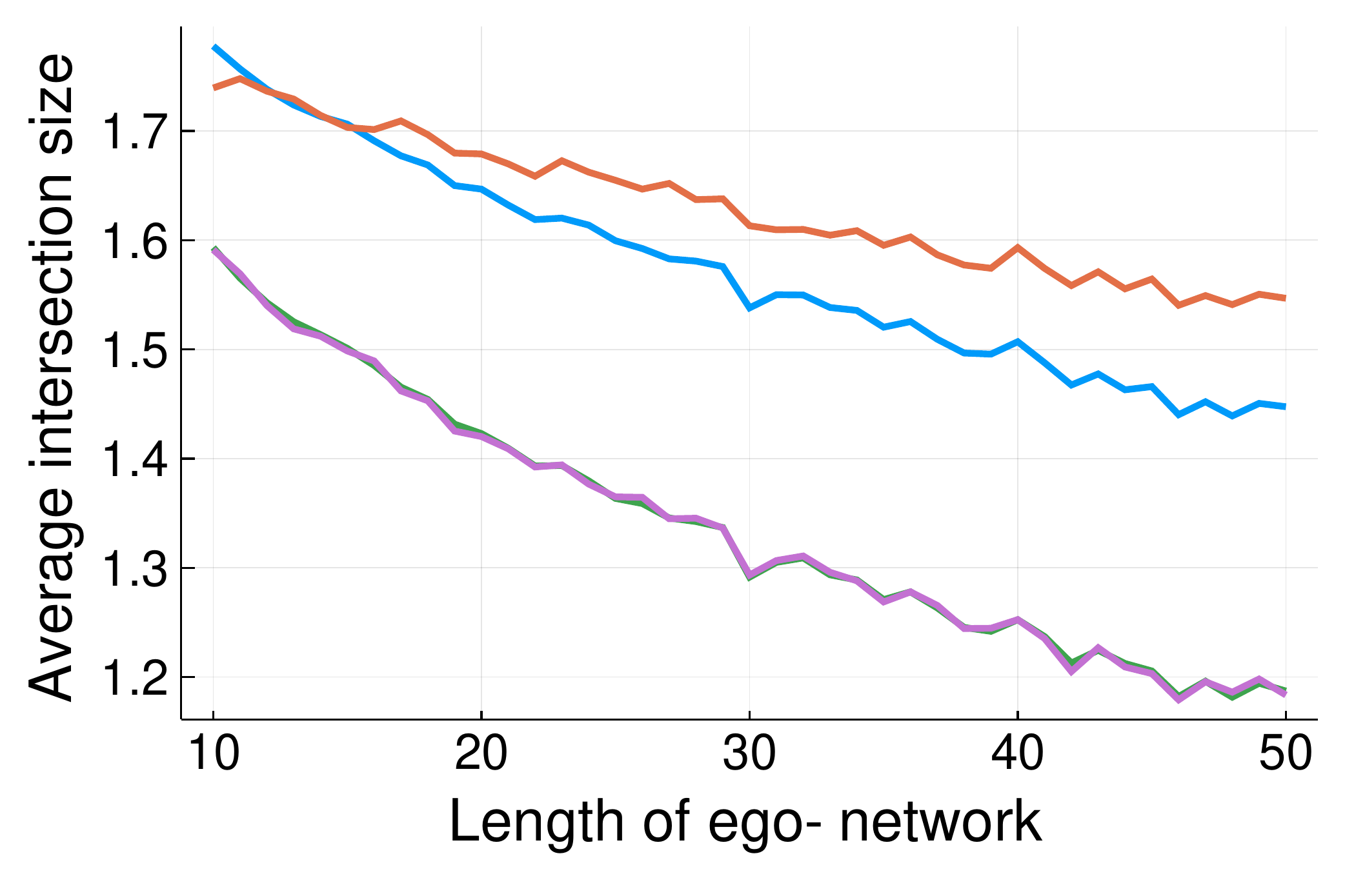}}
\subfigure[Star - email-Avocado]{\includegraphics[width=6cm]{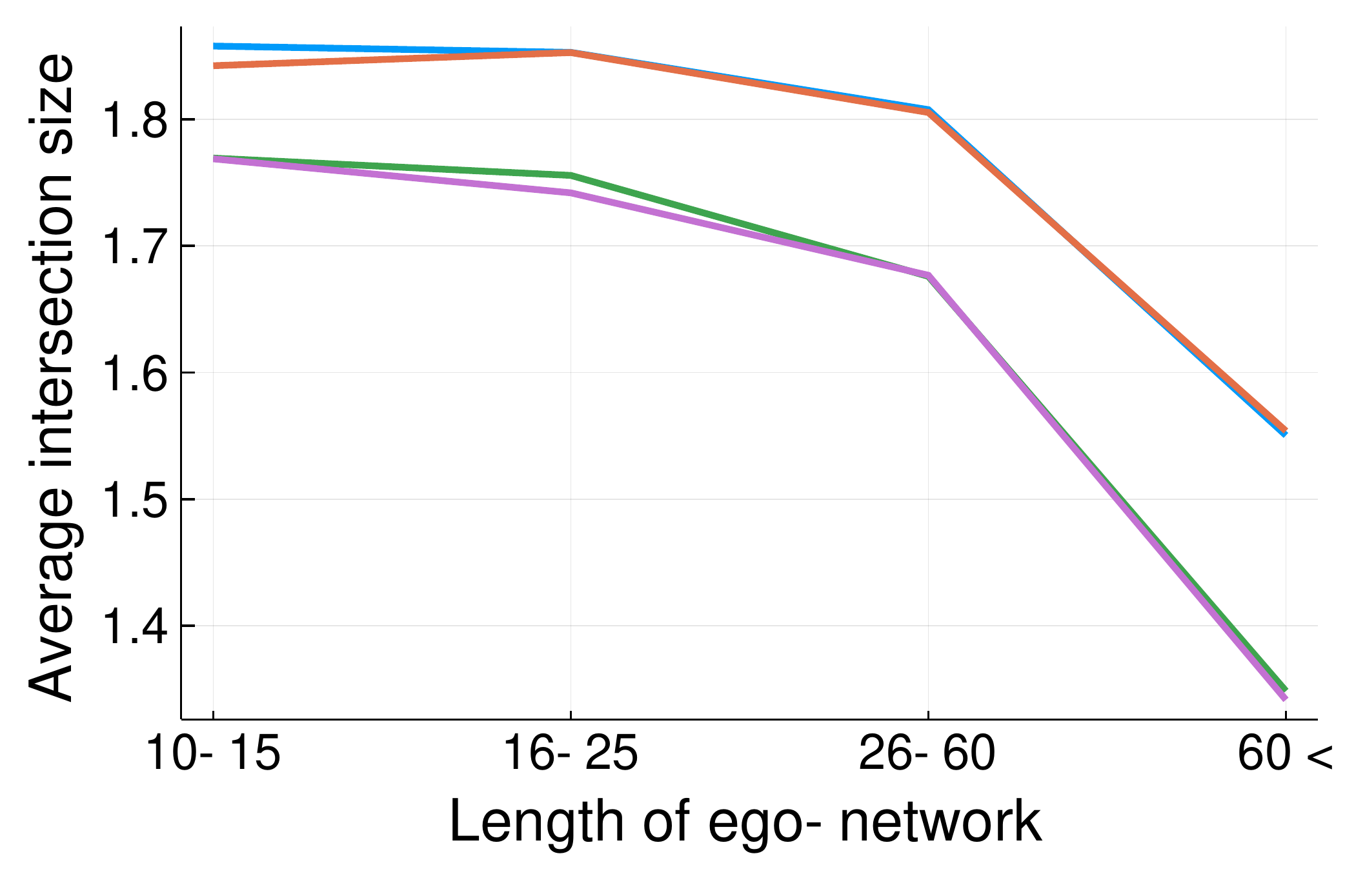}}
\subfigure[Radial - email-Avocado]{\includegraphics[width=6cm]{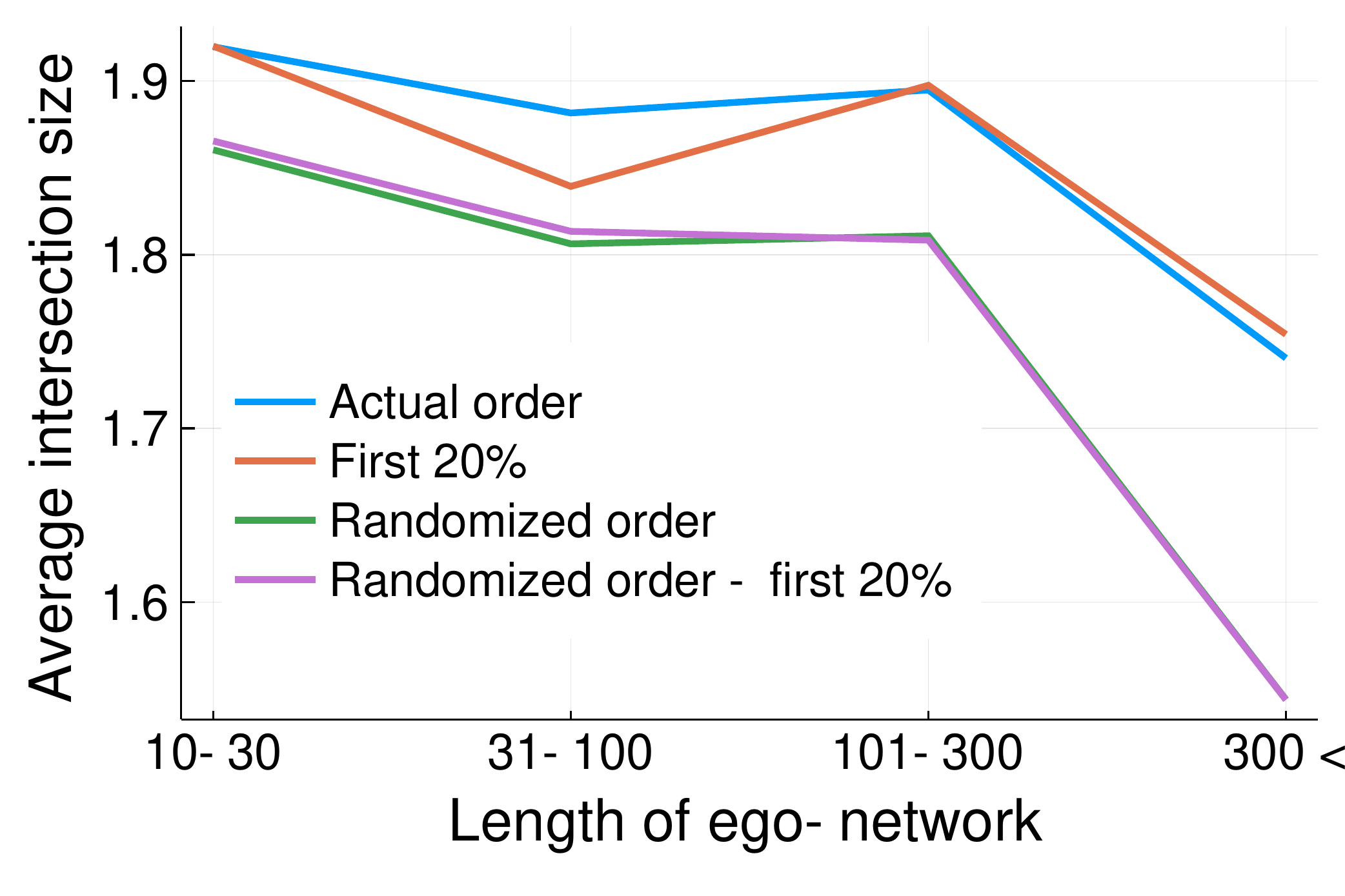}}
\subfigure[Contracted - email-Avocado]{\includegraphics[width=6cm]{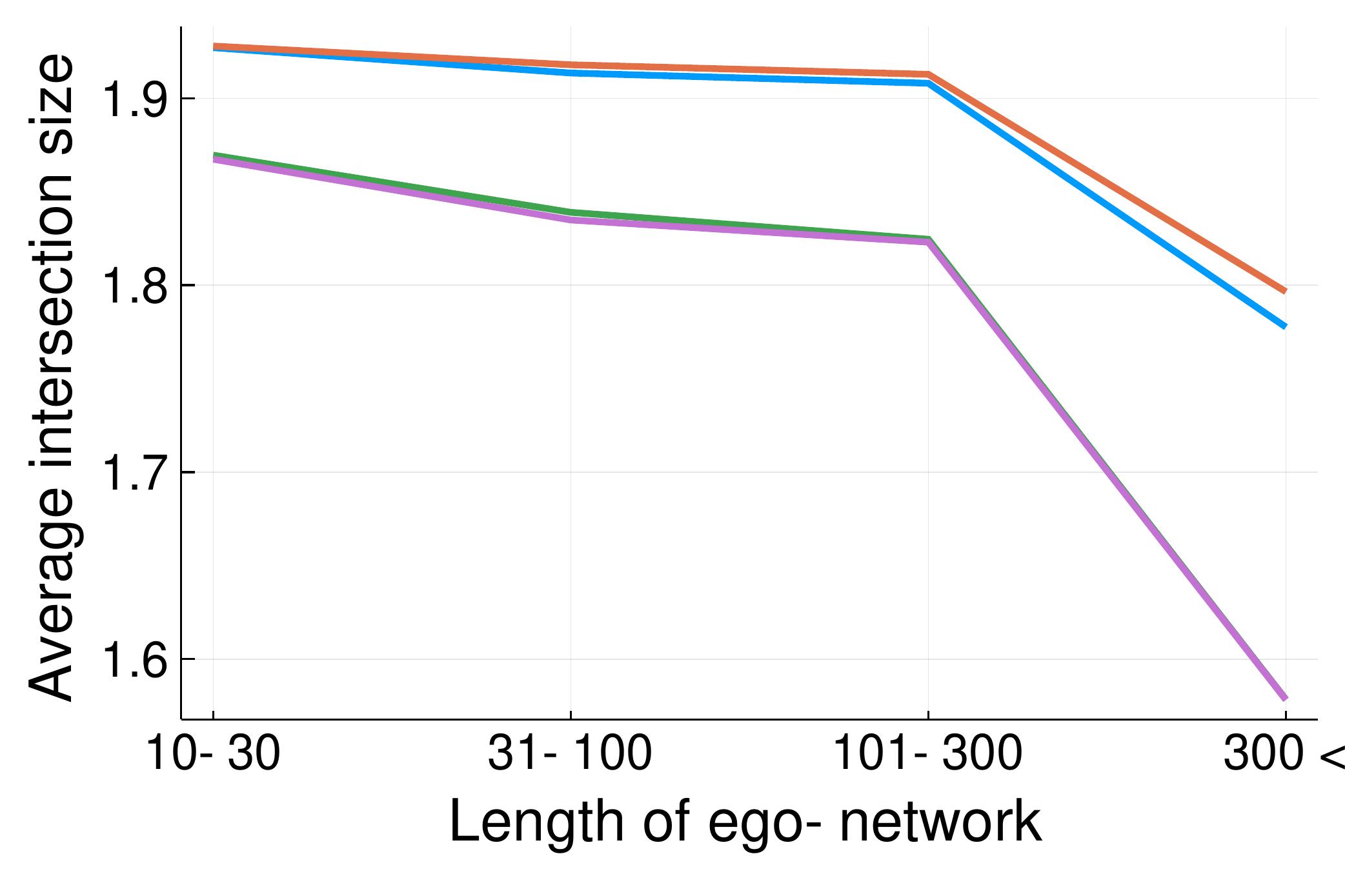}}
\caption{\textbf{Average intersection size. From the plots, we observe that adjacent simplices in correctly ordered ego-networks across all types and domains intersect each other far more an average that adjacent simplices in randomly shuffled ego-networks. This key finding will be significant for our prediction task, acting as a key differentiator between correctly ordered and randomly ordered ego-networks.}}
\label{fig:int_size}
\end{figure*}

\section{Datasets}
\noindent We proceed by describing the three datasets used in this paper. Each dataset consists of a set of timestamped simplices, and from this set, we are able to build star, radial, and contracted ego-networks for each user node.  We collected data from three different domains, emails, online threads, and publications. The three datasets used are useful to study for this problem as they contain entire lifetimes of user activity. For example, in email-Avocado, every email that a given ego has ever been a recipient of or has sent has been recorded from the first to the last. This allows us to accurately analyze the temporal structure of each user’s ego-network. 
Our code uses the ScHoLP library to extract ego-network information from higher-order datasets. 
\par

\begin{itemize}
\item \textbf{coauth-DBLP: 1569217 nodes}. Each node is a researcher, and each simplex corresponds to a set of authors on a publication. The timestamp of each simplex corresponds to the year the paper was published.\par
\item \textbf{email-Avocado: 28244 nodes}. Each node is an email address, and each simplex corresponds to a set of recipients and the sender of an email. The timestamp of each simplex corresponds to the second the email was sent. \par
\item \textbf{threads-ask-ubuntu: 33853 nodes}. Each node is a user, and each simplex corresponds to a set of users answering a question on a forum. The timestamp of each simplex corresponds to the second the question was posted. \par
\end{itemize}

It is important to note that for a given ego-network in coauth-DBLP, multiple simplices can exist at the same timestamp. This does not happen in email-Avocado and threads-ask-ubuntu because of the granularity of data collection mentioned above. We deal with this in two ways. Firstly, we convert the timestamps for each simplex from real time to \emph{ordinal time}, which is a consecutive ordering of the simplices as $1, 2, 3 ...$ in order of their arrival times. Then, for each set of simplices that have the same ordinal time label, we add each simplex iteratively into the ego-network as illustrated in Figure \ref{fig:first-example}. \par
In order to capture enough higher-order interaction for our prediction task, we will only examine ego-networks with a length of at least 20 or greater for coauth-DBLP and 10 or greater for email-Avocado and threads-ask-ubuntu. We will only be analyzing non-trivial simplices in each ego-network. Additionally, simplices can never be removed from an ego-network. We also will ignore all ego-networks where the majority of simplices are identical and ego-networks that have less than 10 alters (except for email-Avocado), as any model that attempts to sort these ego-networks will do well. \par

\section{Key Observations And Measures}
\noindent In this section, we examine a sequence of key observations that we identify about the temporal growth of ego-networks in the domains we analyze.  First, we discuss an underlying locality principle, that the same nodes tend to reappear in neighboring simplices in the temporal order. We next observe that each ego-network can be thought of as a union of star-shaped sub-networks, one for each alter, and that analyzing the temporal structure these sub-networks can give us insights into the overall evolution of the ego-network. We demonstrate a relationship between the time at which a user node arrives in their own radial and contracted ego-network and the size of their ego-network. Then, we highlight the typical placement of high-degree nodes in ego-networks. Finally, we discuss how nodes that have never been seen in an ego-network tend to enter the ego-network at a near-constant rate. For the sake of brevity, in this section we will only analyze ego-networks found in coauth-DBLP and email-Avocado (the observations we will discuss still hold for threads-ask-ubuntu), but we will use all three datasets in our prediction task.

\begin{figure*}[]
\subfigure[Star - coauth-DBLP]{\includegraphics[width=6cm]{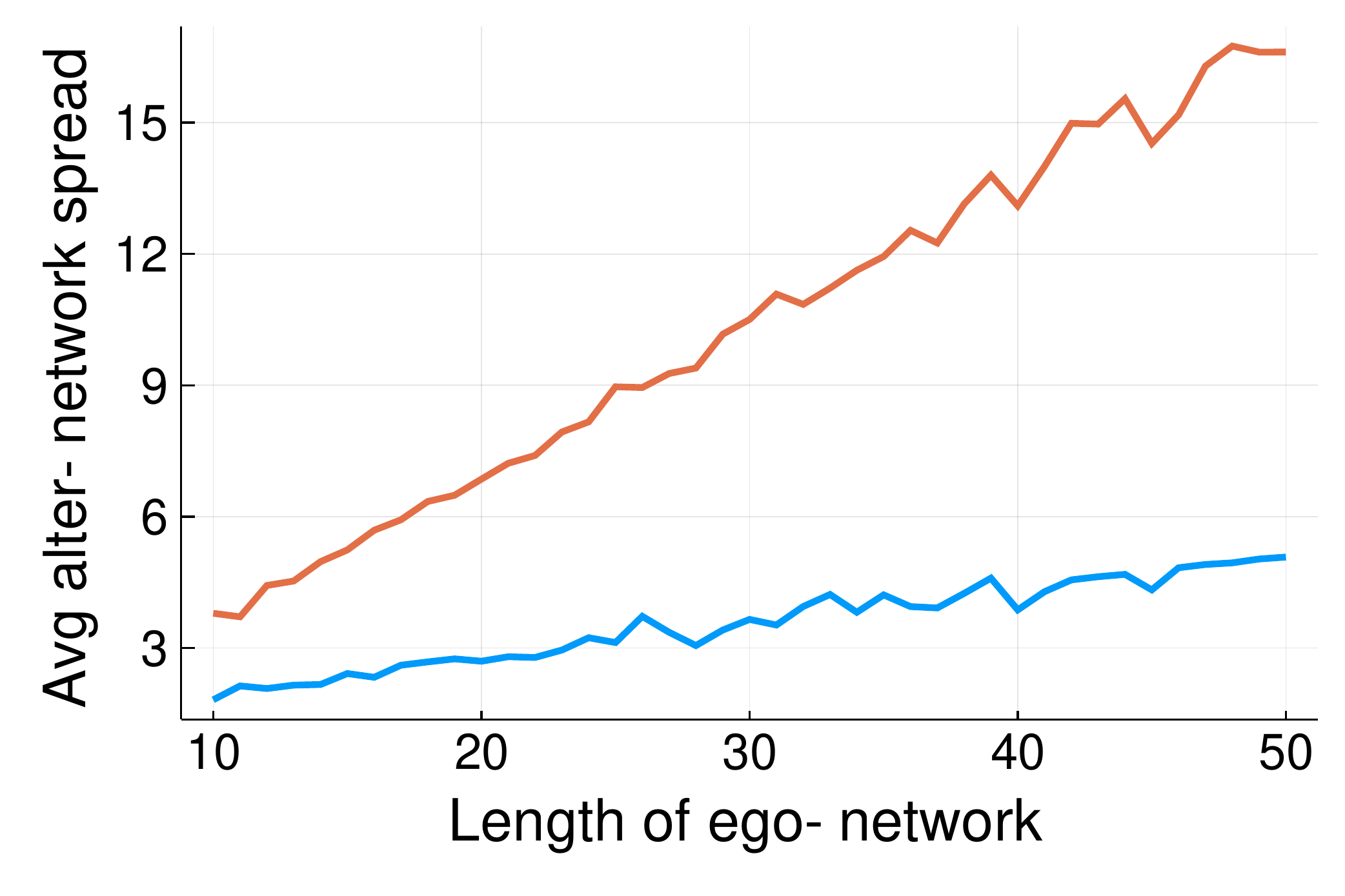}}
\subfigure[Radial - coauth-DBLP]{\includegraphics[width=6cm]{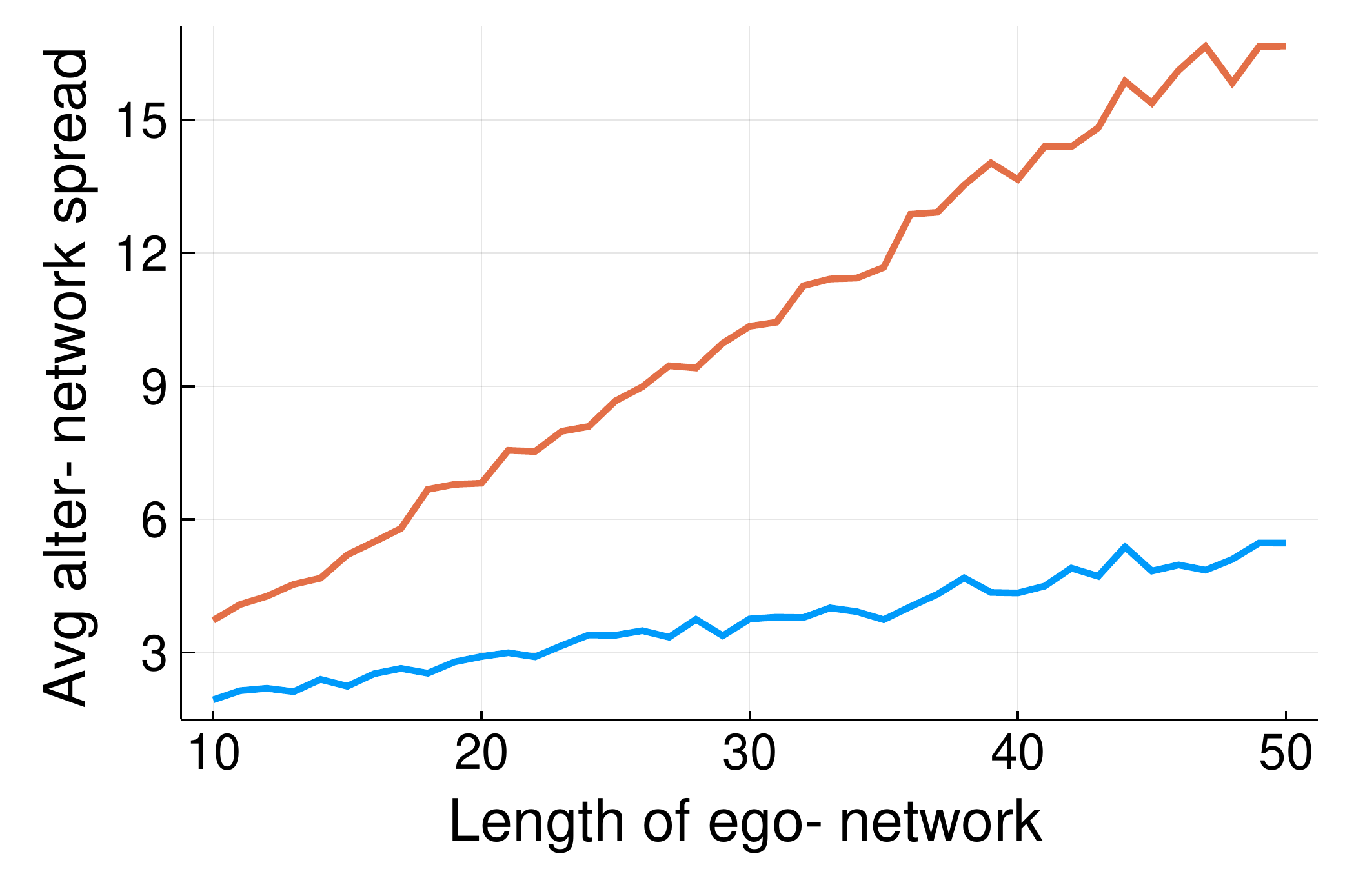}}
\subfigure[Contracted - coauth-DBLP]{\includegraphics[width=6cm]{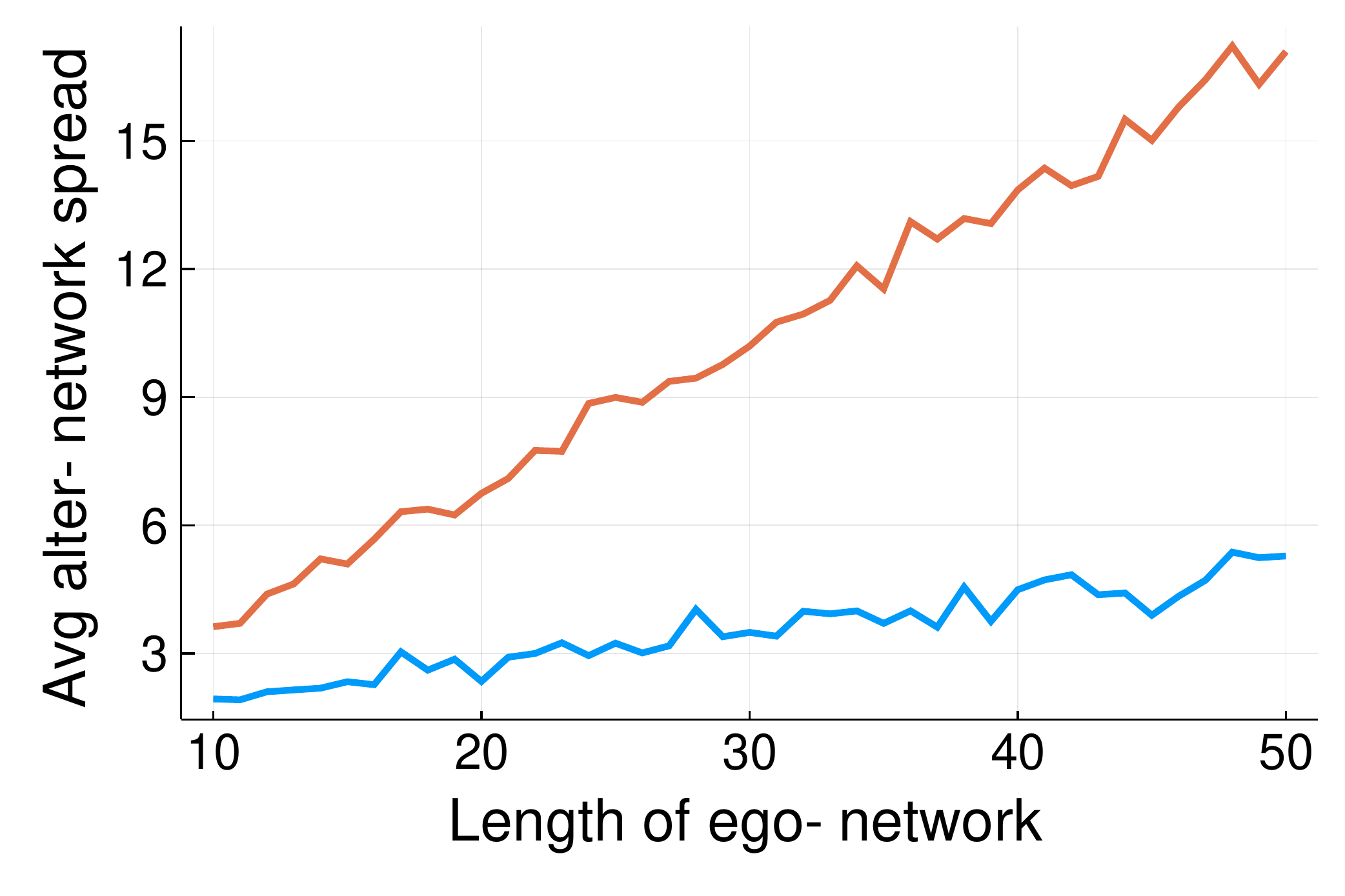}}
\subfigure[Star - email-Avocado]{\includegraphics[width=6cm]{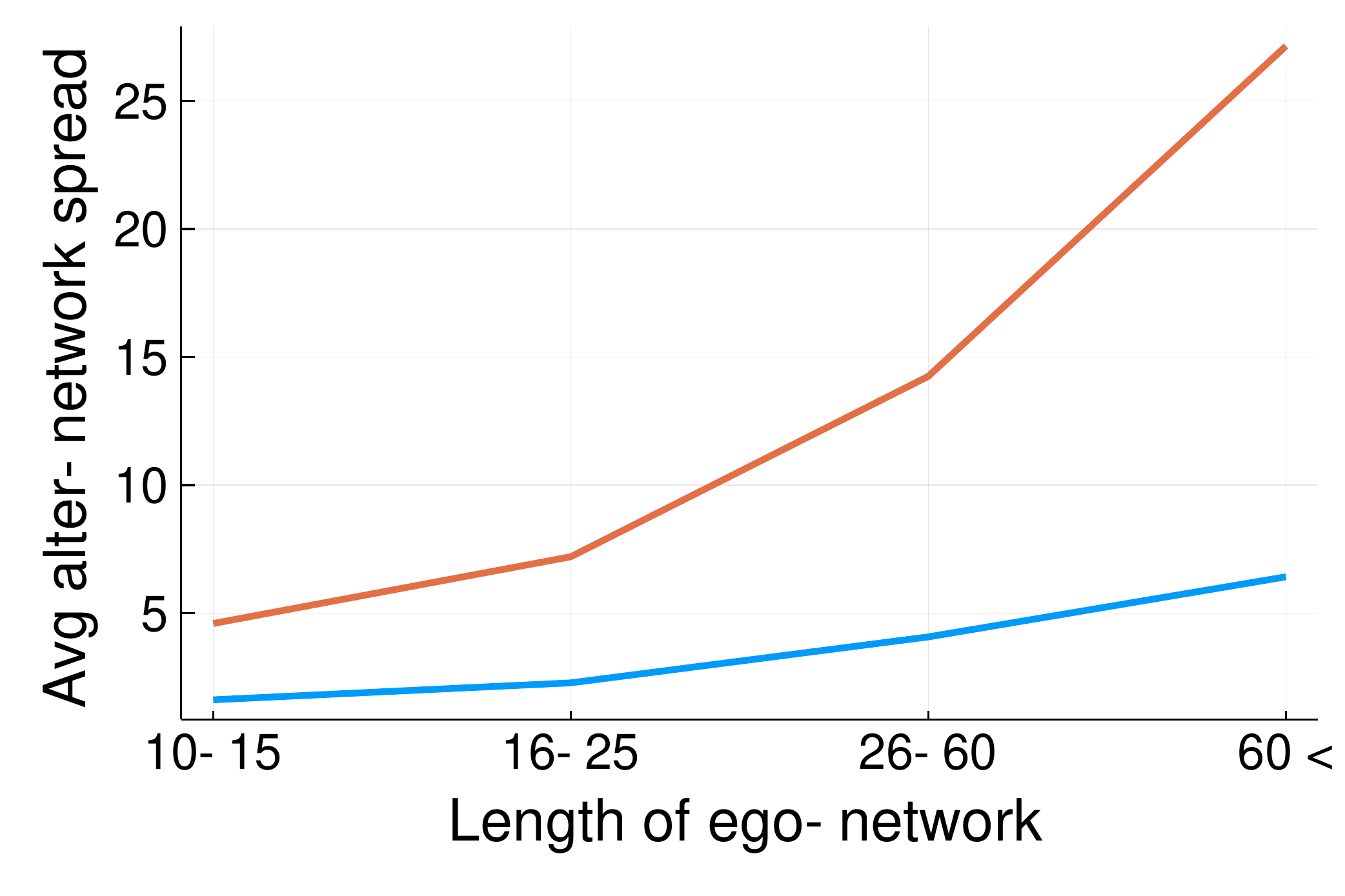}}
\subfigure[Radial - email-Avocado]{\includegraphics[width=6cm]{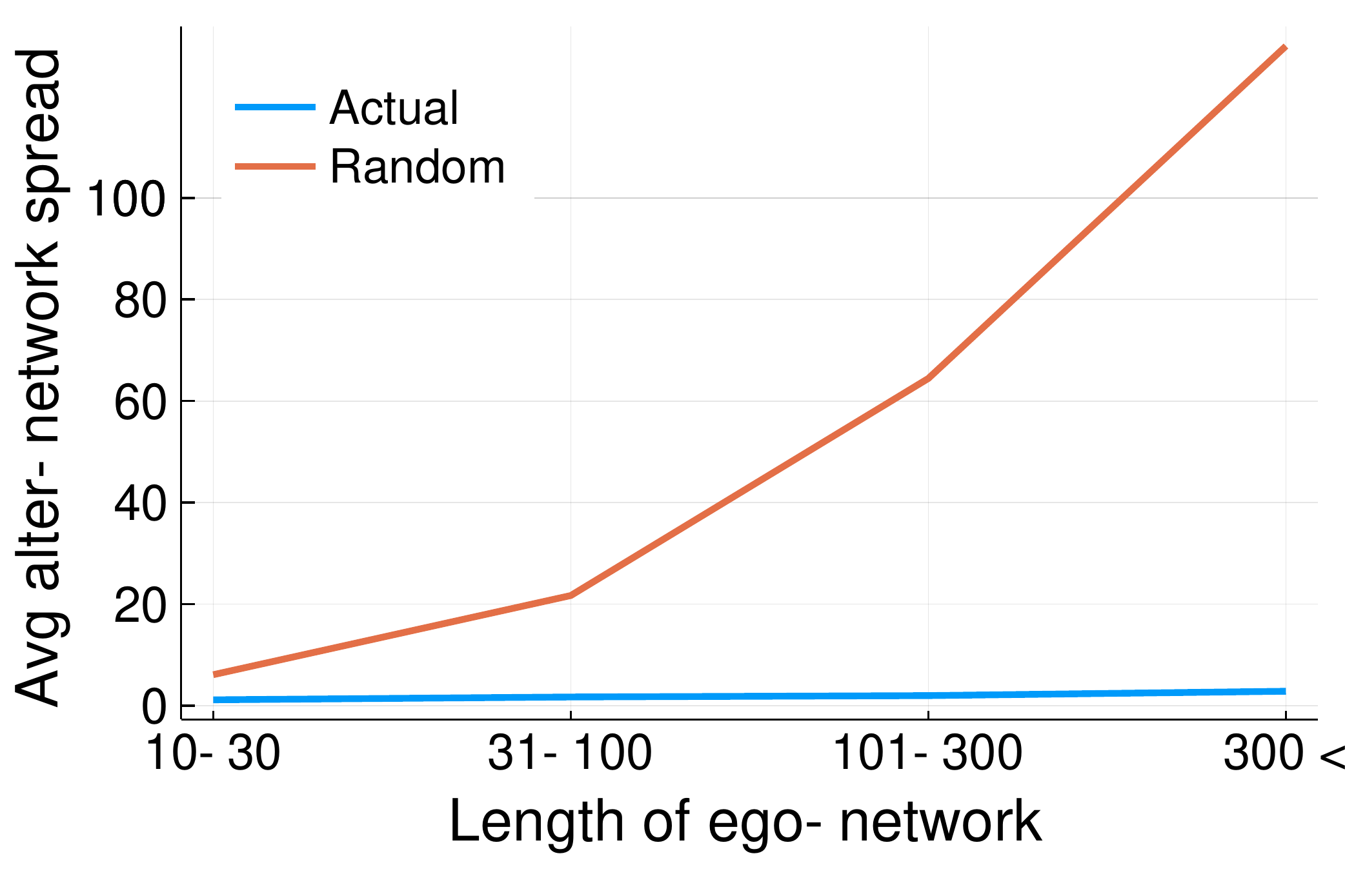}}
\subfigure[Contracted - email-Avocado]{\includegraphics[width=6cm]{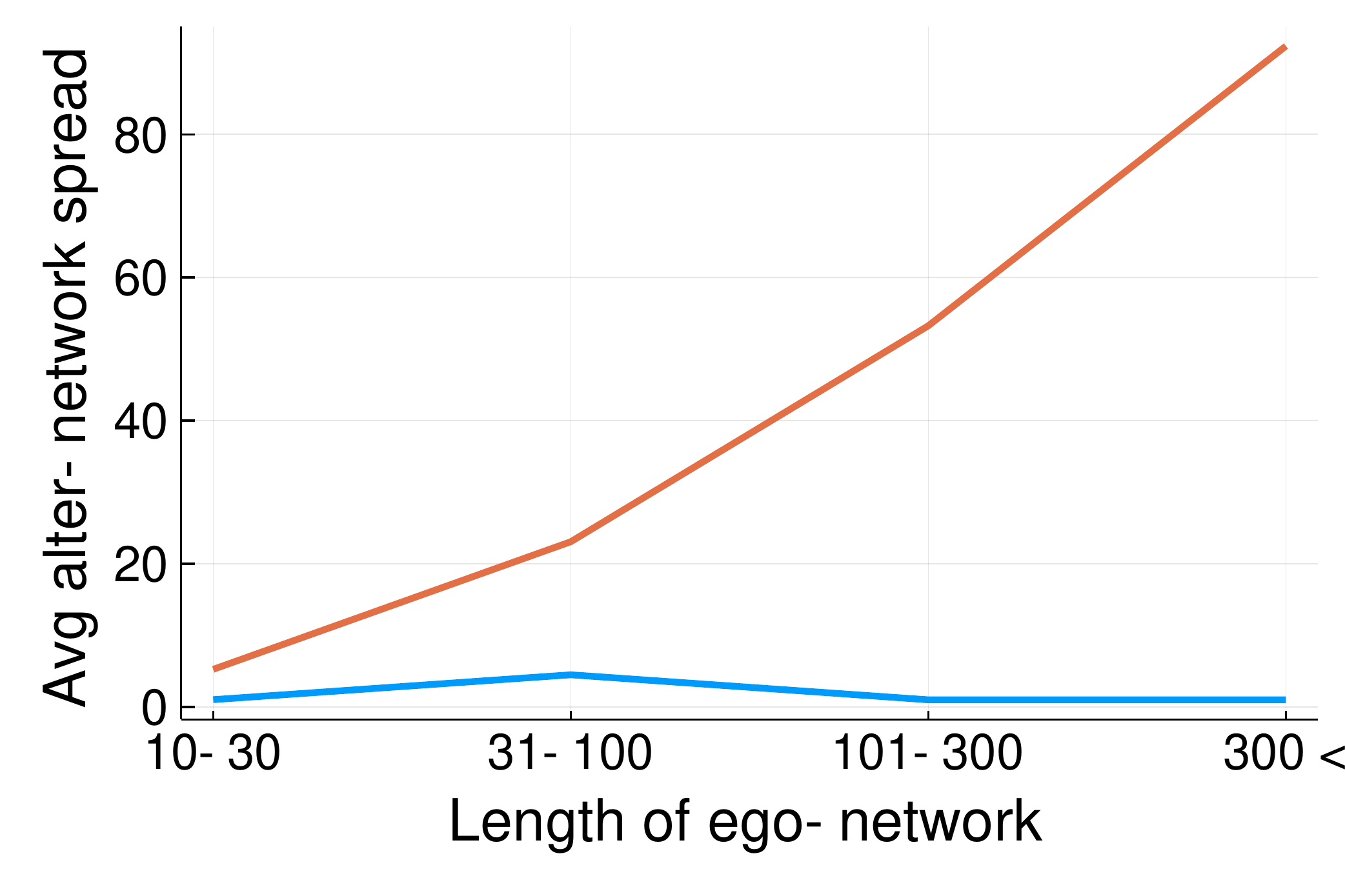}}
\caption{\textbf{Average alter-network spread for all three ego-networks in coauth-DBLP and email-Avocado. We find that, on average, simplices in alter-networks are temporally localized, yet this is not the case if we randomly order the ego-network that each alter-network is a part of. This finding supports the importance of the alter-network spread feature used in our binary classification method.}}
\label{fig:alternet_spread}
\end{figure*}

\subsection{Intersection Size}
\noindent We first observe that contiguous simplices in an ego-network tend to have relatively large intersections with each other, suggesting that temporally adjacent higher-order interactions have high similarity. In other words, the same nodes tend to appear in neighboring simplices. We define the average intersection size of a given ego-network $E$ with $m$ simplices $\{s_1, ..., s_m\}$ as
\begin{displaymath}I(E) = \frac{\sum^{m-1}_{i=1} |s_i \cap s_{i+1}|}{m-1} \end{displaymath}
Figure \ref{fig:int_size} shows the average intersection size of all star, radial, and contracted ego-networks against the size of the ego-network for each dataset. We also plot the average intersection size in the first 20\% of each ego-network and the average intersection size of randomly shuffled ego-networks. \par
From Figure \ref{fig:int_size}, we first observe that adjacent simplices in randomly shuffled ego-networks do not intersect as much as correctly ordered ego-networks. This remains true across all ego-network types in both coauth-DBLP and email-Avocado. Therefore, higher-order interactions in real-world ego-networks tend to be similar to temporally adjacent interactions. Secondly, we find that in coauth-DBLP, the first few simplices in both correctly-ordered star ego-networks and large radial and contracted ego-networks are highly similar, with large intersection size across many pairs of adjacent simplices. However, this is not the case with randomly shuffled ego-networks or with small radial and contracted ego-networks, where there is no significant difference between the average intersection size of the first 20\% of simplicies of a shuffled ego-network compared to the entire ego-network. These observations are consistent with prior work studying sequences of higher-order interactions \cite{benson2018sequences}. \par
A key idea here is that if we observe the average intersection size between adjacent simplices in a traditional pairwise graph, it becomes difficult to assess how similar two adjacent interactions involving more than two nodes are, as both interactions must be broken down into several pairs of nodes. By representing these interactions with hypergraphs, we are able to better capture ideas of intersecting edges. \par
Prior work studying the evolution of dyadic ego-networks finds that ego-networks tend to expand a great deal towards the beginning of their lifetime \cite{aiello2017evolution}. Our findings show that in the higher-order case, the opposite is true for coauth-DBLP, where the first few interactions in a hypergraph ego-network's lifetime are highly similar, with many of the same nodes reappearing in contiguous simplices. These results are particularly significant as we also find that the average size of an incoming simplex for an ego-network increases over time (the details of which we have omitted), implying that despite the fact that these earlier simplices are relatively small, they still intersect a great deal. We conclude that both observations mentioned above are important concepts for any model that attempts to capture the temporal structure of ego-networks. \par

\subsection{Alter-networks}
\noindent We now analyze the temporal spacing between similar simplices in ego-networks and extend some of the ideas previously seen when discussing intersection density. Every alter $a$ of a user $u$ has their own star-shaped collection of all simplices in $u$’s ego-network $E(u)$ that include $a$. We define this set of simplices as $a$’s \textit{alter-network}, $Alt(a, E(u))$. The alter-network of $a$ can also be thought of as the intersection of $a$’s star ego-network with $u$’s ego-network:
\begin{displaymath}Alt(a, E(u)) = E(u) \cap T(a) = \{s: (a \in s)\}, \forall s \in S \end{displaymath}
where $S$ is the set of all simplices in $E(u)$. Every ego-network of any type can be thought of as a union of its alter’s interactions. Stated differently, for the ego-network $E(u)$ of a user node $u$ and the set of alters $A(u) = {a_1, ..., a_n}$:
\begin{displaymath}E(u) = \bigcup_{i=1}^{n} Alt(a_{i}, E(u))\end{displaymath}
Knowledge of the temporal structure of the alter-networks in a given ego-network proves to be very useful when predicting the evolution of the ego-network. We define the \textit{alter-network spread} of a given alter-network to be the average ordinal time difference between two adjacent simplices in the alter-network. Stated differently, let $t(s)$ be the timestamp at which simplex $s$ arrives in the ego-network $E(u)$ of some user $u$. Then, the alter-network spread of an alter-network of size $m$ composed of simplices $\{s_1, ..., s_m\}$ is:
\begin{displaymath}\frac{t(s_m) - t(s_1)}{m-1} \end{displaymath}
For each alter-network in coauth-DBLP and email-Avocado, we measure the average alter-network spread and plot the average value against the size of the ego-network, as shown in Figure \ref{fig:alternet_spread}. We compare against a random model that calculates the average alter-network spread for a shuffled alter-network. From this, we observe that the spacing between neighboring simplices in a given alter-network is relatively small. This is not the case for randomized alter-networks, which have a higher average alter-network spread. This proves to be a vital feature for our learning model, as the existence of dense alter-networks for a given ego-network is strong evidence that the ego-network is correctly sorted. \par
The intuition behind the alter-network spread is that it captures the period of time in which a user is interacting with a particular alter, which above we confirm to be temporally local. However, an interesting question to ask is whether or not the rate at which users' interact with their alters is constant. Using the example of co-authorship, it is intuitive to think that the first and final few papers a user co-authors with a frequent collaborator are less temporally consistent than the papers they publish in between. It is natural to conjecture that this middle section is the most dense, and therefore would have the smallest alter-network spread. For large alter-networks, which we will define to be of size at least 10, we also measure spread at different sections of the alter-network. To do this, we split alter-networks into thirds, and calculated the average alter-network spread in each third. We observe that the middle third of large alter-networks are in fact much denser on average than the beginning and final third. Thus, on large alter-networks, our model should capture this pattern. \par
When comparing average alter-network spread to average intersection size, these two features are in no sense identical, but they are certainly related. We would expect to find an ego-network with high average intersection size to have low alter-network spread, as intersecting adjacent simplices implies adjacent simplices in the alter-networks of alters that are common across both simplices. This relationship can be seen by comparing Figures \ref{fig:int_size} and \ref{fig:alternet_spread}. As the length of ego-networks increase, the average intersection size decreases while the average alter-network spread increases. \par
Finally, we observe that high-degree alters on average tend to appear earlier on in ego-networks than low-degree alters (we omit the details in this paper). In the context of co-authorships, this observations means that if $w$ is one of some user node $u$'s most frequent co-authors, then edges containing $w$ are more likely to appear earlier than later. This finding is intuitive and especially important for models attempting to solve the problem of temporal reconstruction, as the first few simplices of any correctly ordered ego-network are likely to contain high-degree nodes.

\subsection{An Anthropic Principle for Ego-Networks}
\noindent In an ego-network, the user node $u$ occupies a privileged position at the center, but many of the higher-order interactions in the radial and contracted ego-networks do not involve $u$, and in fact might have pre-dated the first interaction that does involve $u$.  
Thus, the ego-network is defined by $u$, but parts of it pre-dated $u$'s arrival into the system.
We refer to this tension as an \emph{anthropic principle for ego-networks}, by analogy with the collection of anthropic principles in physics and philosophy \cite{carter1974large} that involve a similar duality: that your position as an observer is privileged, but that the system you are observing existed before you were there to observe it.
\par
We now consider these ideas in more detail. As an example in terms of co-authorship, two alters in some user node $u$’s ego-network may have written papers together before ever meeting $u$ (Figure \ref{fig:first-example}(c)). This is only the case for radial and contracted ego-networks, as the star ego-network does not include interactions among alters that omit $u$. In a star ego-network, the user node will always arrive at the very first timestamp. In contrast, the radial and contracted ego-network of a particular user node may have begun a long time before the user node ever entered it.\par
\begin{figure}

\includegraphics[width=7cm]{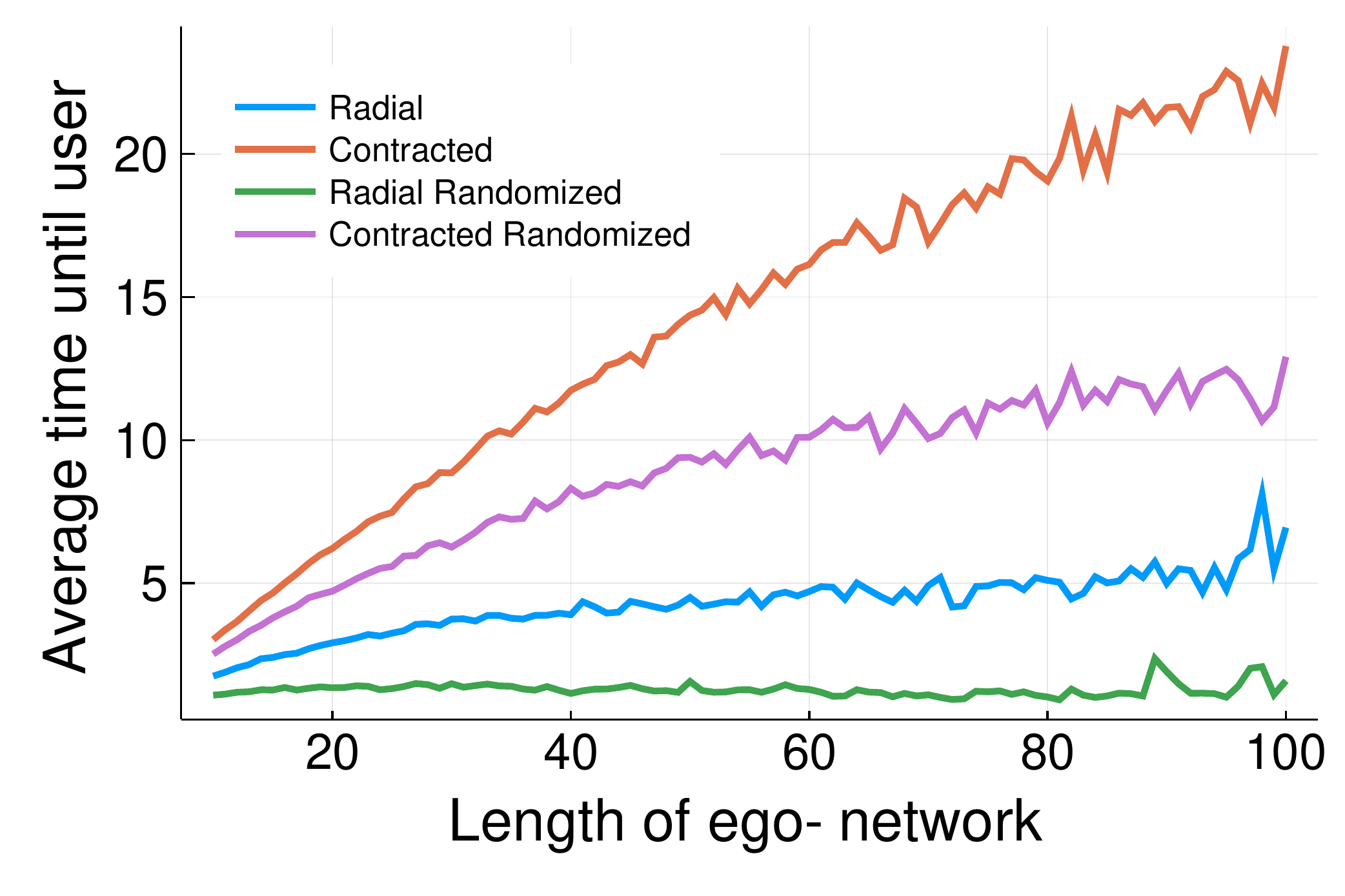}

\caption{\textbf{For coauth-DBLP, the user node tends to arrive into its radial ego-network before or around the fifth timestamp on average. For contracted ego-networks, there is a linear relationship between the length of the ego-network and the timestamp at which the user node arrives in the ego-network. This becomes a key feature for our classifier on radial and contracted ego-networks in coauth-DBLP.}}
\label{fig:time}
\end{figure}

An interesting question to then consider is: across real-world radial and contracted ego-networks, at what timestamp does the user node typically arrive in their own ego-network? For coauth-DBLP, we plot the average timestamp at which a user node arrived in their correctly ordered and randomly shuffled radial and contracted ego-networks against the size of the ego-network, as shown in Figure \ref{fig:time}. Here, we observe a linear relationship between the timestamp at which a user node arrives and the size of their contracted ego-network. User nodes in randomly shuffled contracted ego-networks tend to arrive far earlier, implying that alter simplices tend to dominate the earlier sections of real-world contracted ego-networks. \par
However, we observe a sublinear relationship for radial ego-networks, as the user node usually enters their ego-network around or before the fifth timestamp. An interesting observation is that the user node tends to arrive almost at the very first simplex in expectation in randomly shuffled radial ego-networks, implying not only that radial ego-networks are largely made up of user simplices, but also that the fraction of simplices in radial ego-network that include the user node is near-constant as the size of the ego-network increases. We conclude that there is an element of predictability when observing the time at which user nodes enter their radial and contracted ego-networks in co-authorship datasets, and this arrival time becomes a useful feature for our learning model. \par

\subsection{Novelty Rate}
\noindent We now measure the rate at which novel nodes enter ego-networks in order to understand when and how often new nodes typically arrive. We define the \textit{novelty} of a simplex to be the number of nodes in the simplex that have never been contained in any previous simplices. For example, in Figure \ref{fig:first-example}(a), simplex $t_5 \rightarrow \{1, 5, 7\}$ has a novelty of 2 since the nodes 5 and 7 do not appear in simplices $t_1, ..., t_4$. \par
\begin{figure*}[]
\subfigure[Star Ego-network]{\includegraphics[width=5.5cm]{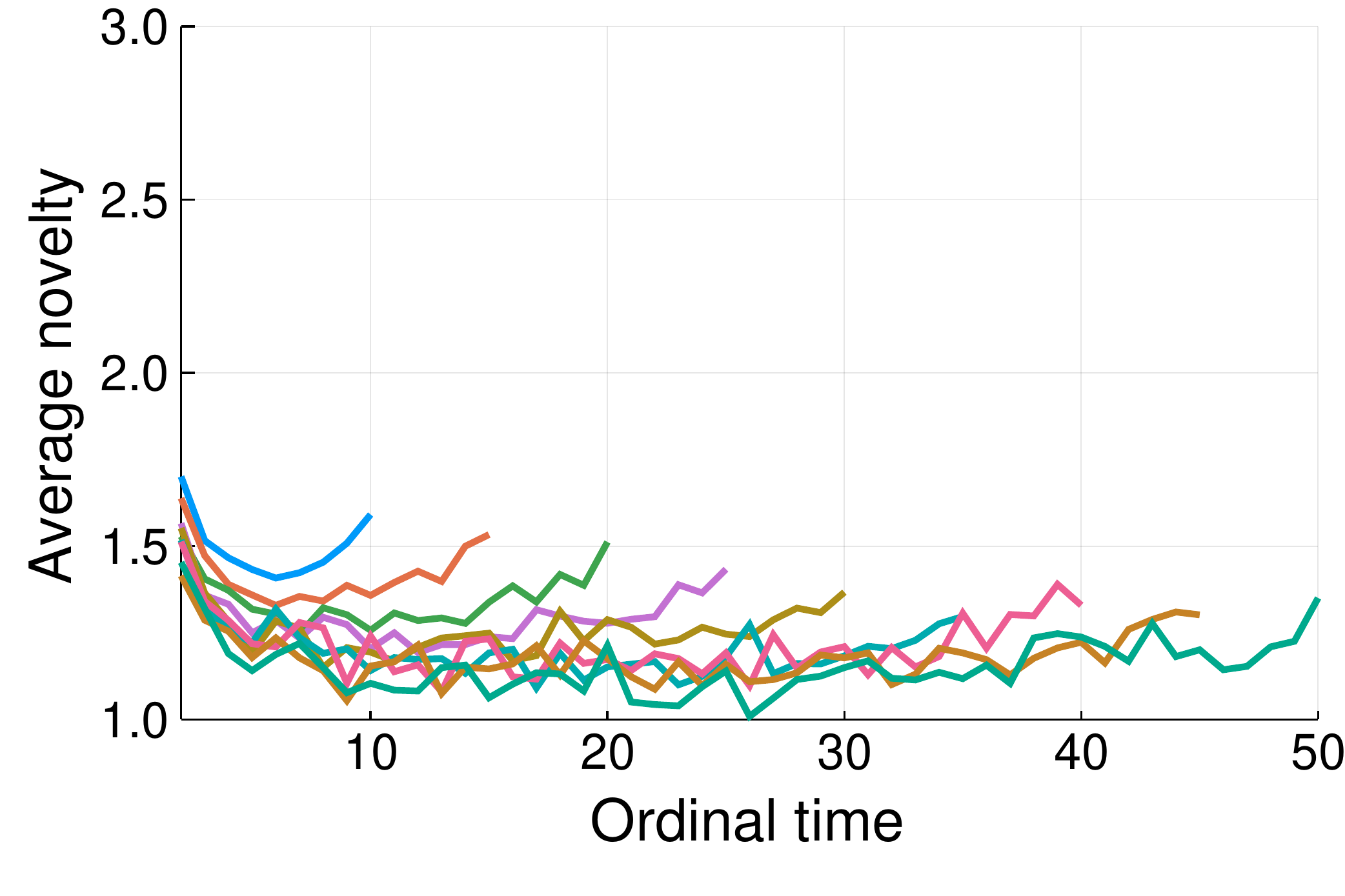}}
\subfigure[Radial Ego-network]{\includegraphics[width=5.5cm]{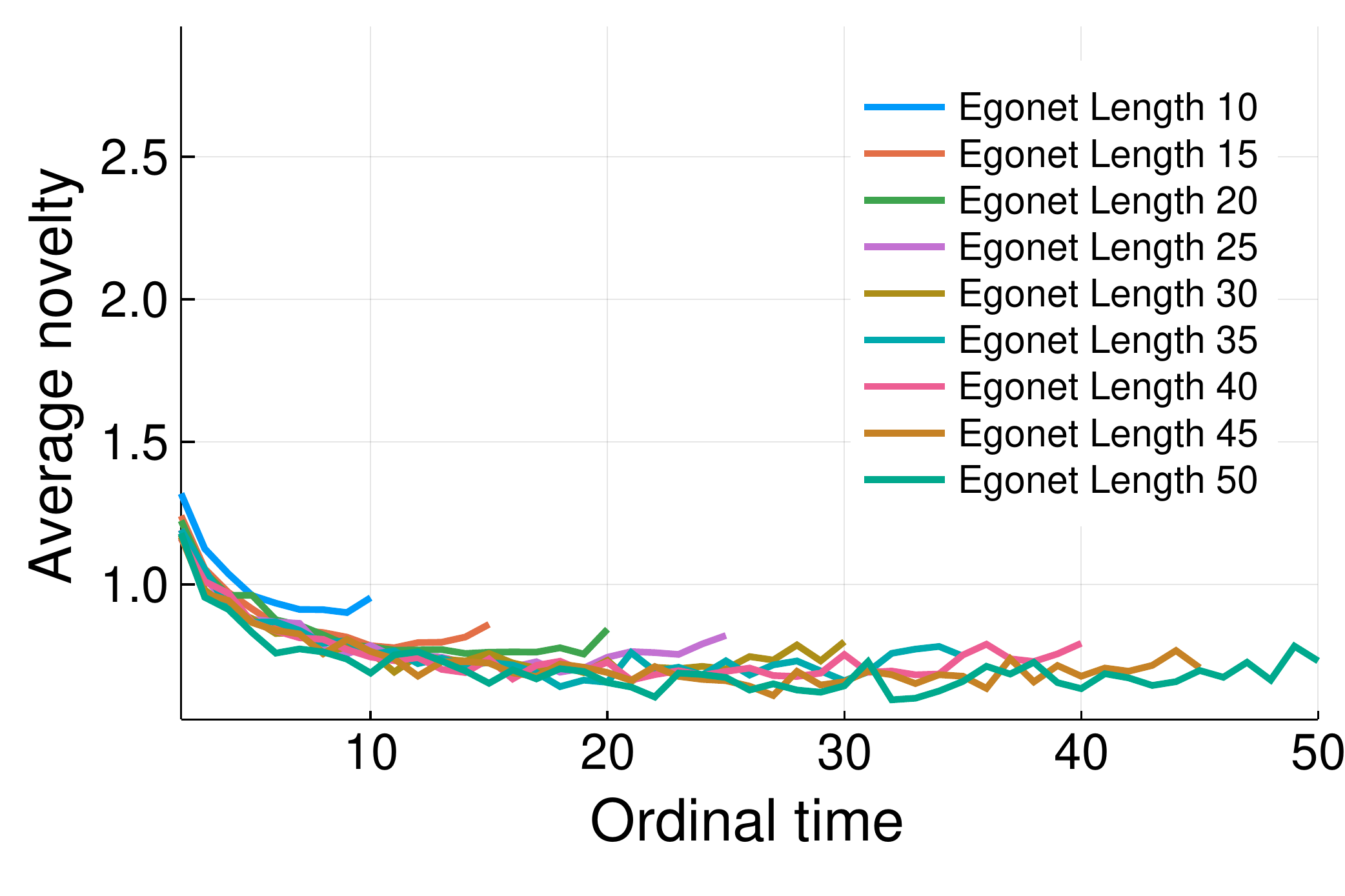}}
\subfigure[Contracted Ego-network]{\includegraphics[width=5.5cm]{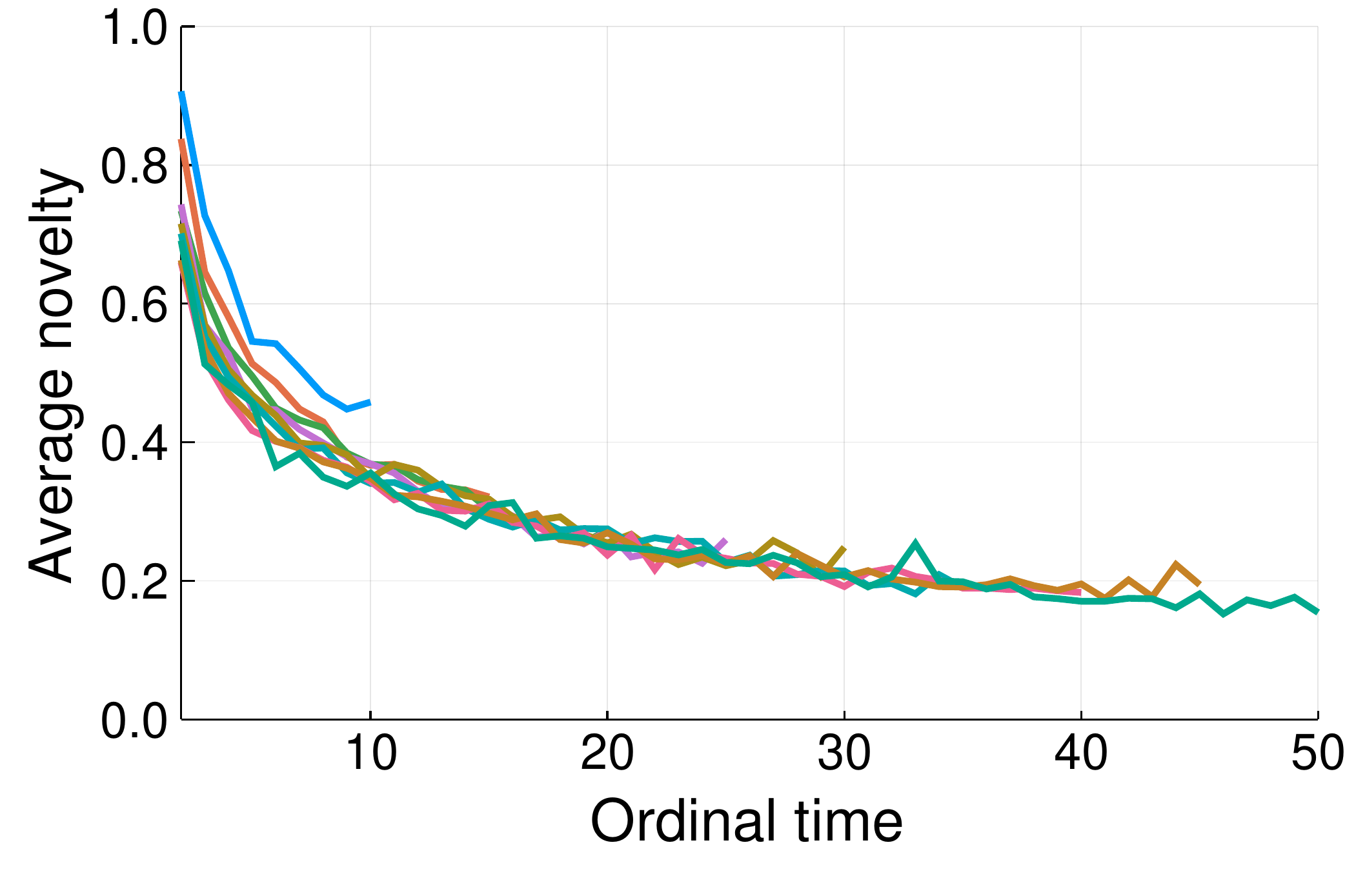}}
\caption{\textbf{Average novelty across the three ego-network types.}}
\label{fig:novelty}
\end{figure*}

Figure \ref{fig:novelty} reports the average novelty at each timestamp for ego-networks of different types and sizes. For each plot, we omit the average novelty of the first simplex in each ego-network. We do this because this value will have novelty equal to the average simplex size of the first simplex, since each node in the simplex is novel. From Figure \ref{fig:novelty}, we observe that for star and radial ego-networks, the number of novel nodes that enter an ego-network at each timestamp slowly and gradually decrease, until a certain point where novel nodes begin to enter the ego-network again. This agrees with our earlier finding that low-degree nodes tend to enter ego-networks later on in the ego-network's lifetime. It is also interesting to note that for star ego-networks, the average novelty of a simplex is always above 1 regardless of the ego-network's size, implying that the average incoming simplex has at least a single novel node. For contracted ego-networks, the rate at which novel nodes enter the ego-network is always decreasing but very slowly. These observations imply that co-authorship ego-networks grow at a near-constant rate for the majority of their lifetimes.

\omt{
\begin{figure*}[]
\subfigure[Star Ego-network]{\includegraphics[width=5cm]{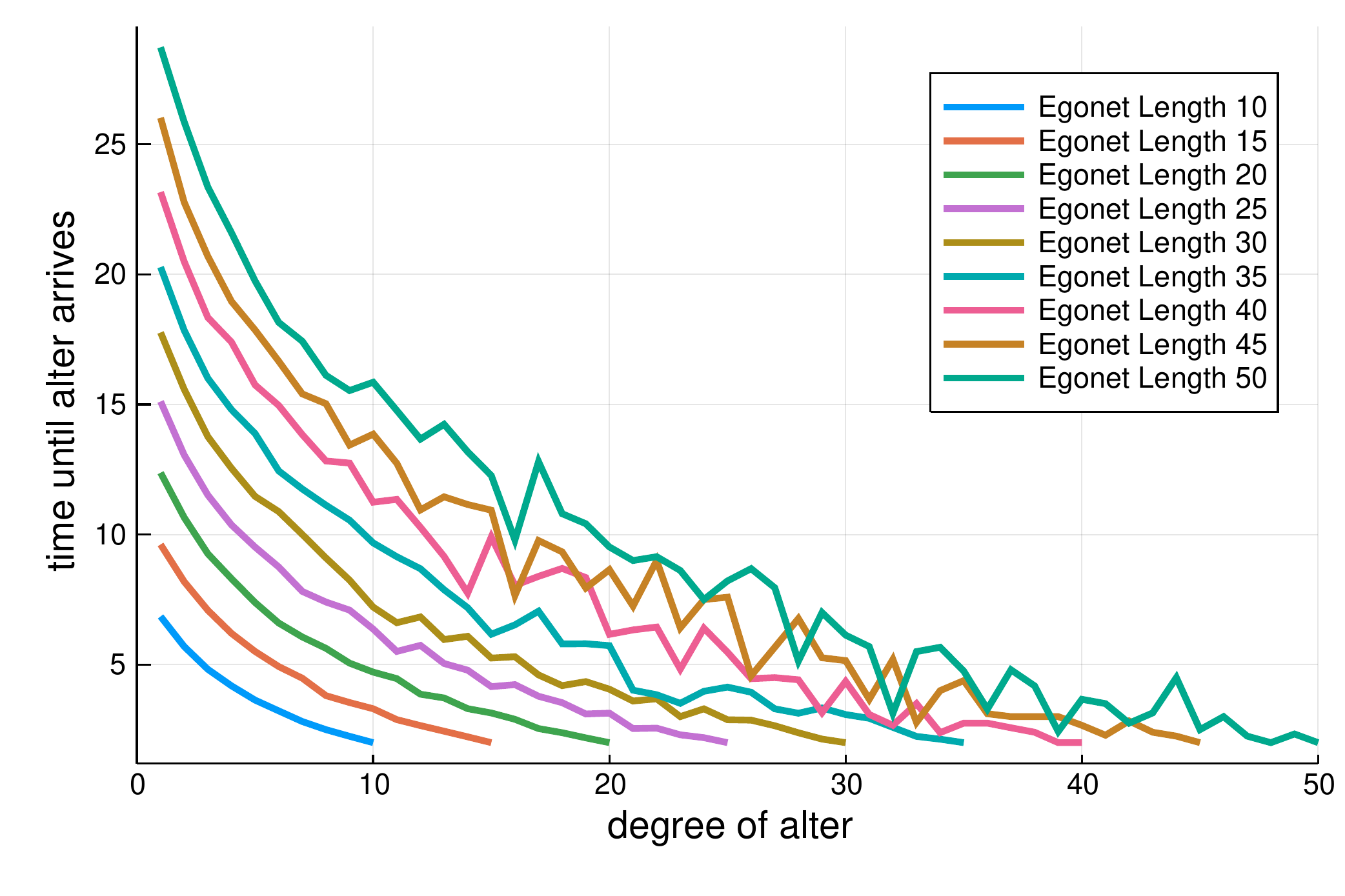}}
\subfigure[Radial Ego-network]{\includegraphics[width=5cm]{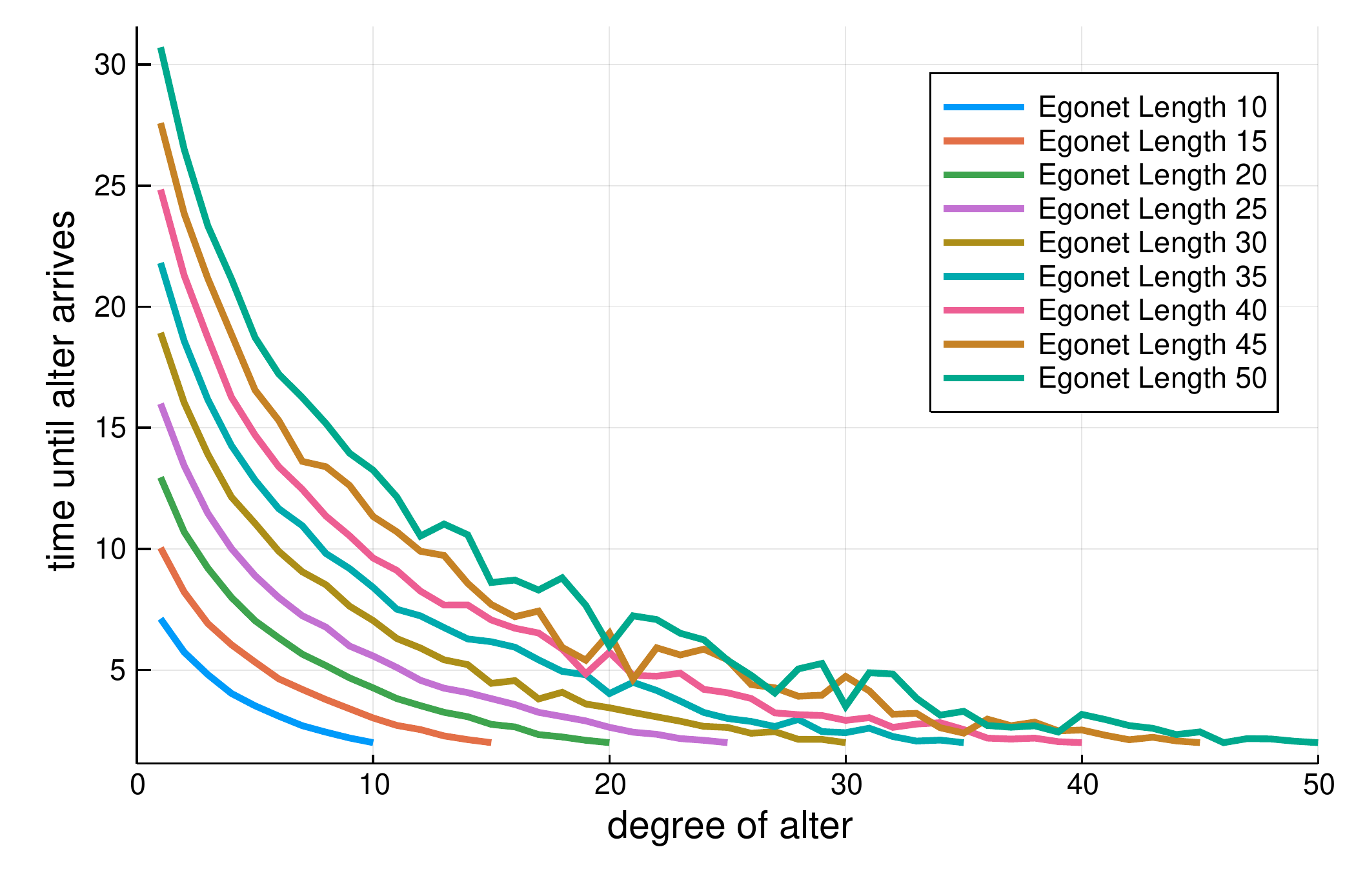}}
\subfigure[Contracted Ego-network]{\includegraphics[width=5cm]{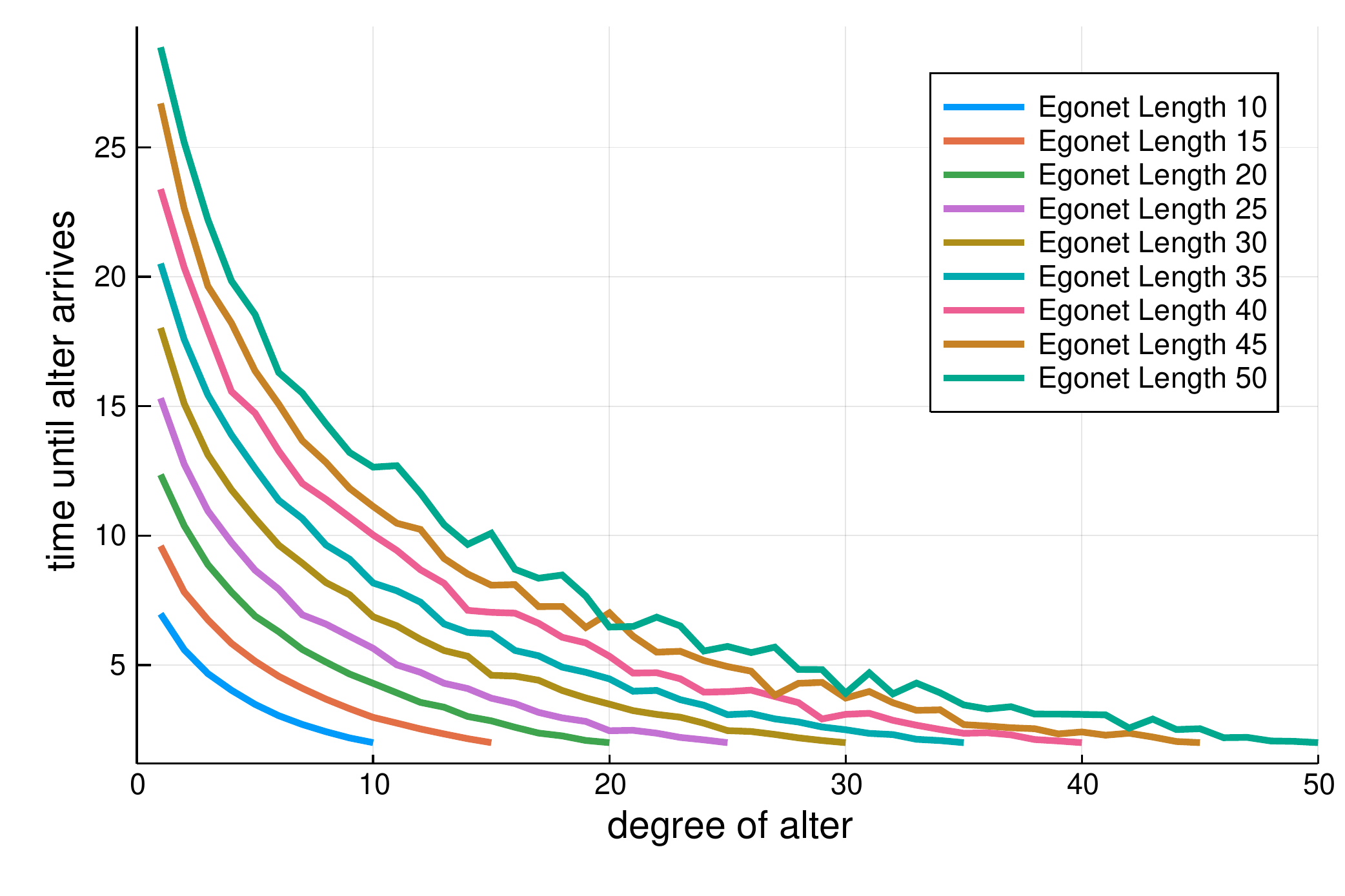}}
\caption{\textbf{Average timestamp at which alters enter star, radial, and contracted ego-networks in coauth-DBLP. We observe that, on average, high-degree alters arrive into ego-networks earlier than low-degree alters for all three ego-network types.}}
\label{fig:time_alter}
\end{figure*}
}
\section{Prediction Task}
\noindent We now discuss the problem of temporal reconstruction of hypergraph ego-networks and detail our solution. 
We are guided by two related goals in formulating and studying this problem: first, for applications where we might have the structure of a hypergraph ego-network but lack information about how it evolved; and second, as a way to evaluate and gain insight into the mechanisms of ego-network evolution in hypergraphs, and how the findings in the previous section might be used for their analysis. Our solution can be broken down into two tasks: first, we will describe the supervised learning task of classifying a given ego-network as correctly sorted or randomly shuffled, and after we will detail a local search algorithm used to iteratively sort a shuffled ego-network.
\subsection{Learning Task}
\noindent We define a supervised binary classification task, where we predict whether a given ego-network is correctly or randomly ordered. Half of our training examples will be correctly sorted ego-networks, and the other half will be shuffled. Therefore, random guessing achieves an accuracy of 50\%. The lengths of radial and contracted ego-networks for email-Avocado and threads-ask-ubuntu are usually large (many of which have length greater than 100), and therefore we will only train our model on star ego-networks. For each ego-network type for coauth-DBLP, we will train and test our model on a sample of 50000 ego-networks of length 20 to 50. For email-Avocado and threads-ask-ubuntu we will sample approximately 1000 and 1800 star ego-networks of length at least 10 respectively.\par
We will construct a set of features from each ego-network. For star ego-networks, these will include features based on the main findings of the previous section: intersection density (the average intersection size of an ego-network divided by average simplex size of the ego-network) and average alter-network spread. We will also use the length of the ego-network, the number of future simplices the first simplex in the ego-network is a subset of, and the number of prior simplices the last simplex is a superset of. According to \cite{benson2018sequences}, early sets in a sequence of sets tend to be subsets of future sets, and later sets tend to be supersets of prior sets. For radial and contracted ego-networks, we will additionally include the timestamp at which the user node entered the ego-network. \par
We trained several classification models on our data, namely deep neural networks, logistic regression, random forests, and naive Bayes. As the neural network performed the best, we will report accuracies from the neural network. We trained three different and fully connected neural networks, one for each dataset. For coauth-DBLP and threads-ask-ubuntu, our network had two hidden layers of sizes 100 and 24, and the network for email-Avocado had two layers of sizes 12 and 6. Each network had an initial learning rate of 0.001 with minibatch size of 200. For each model, we performed 10-fold cross validation and will report the mean classification accuracy. All models were trained using the sci-kit learn library. Training and testing each model took less than a minute using a Lambda 72-core GPU server (1536 GB RAM). \par
Table \ref{table:ml} reports the classification accuracy of our models. In general, our models perform well across all datasets and ego-network types. Due to the limited number of ego-networks in email-Avocado and threads-ask-ubuntu to sample from, models trained on these datasets report high standard deviation values. When comparing our method to a random guessing baseline, we significantly outperform this baseline on all datasets and ego-network types.
\par
We now quantitatively analyze feature importance across datasets. For all datasets, we find that alter-network spread is by far the most significant feature, achieving accuracies of 89\%, 84\%, and 75\% when used as a single feature across star, radial, and contracted ego-networks respectively in coauth-DBLP. Intersection density is also a significant feature but less powerful.
Finally, we also decided to test the prediction model derived from star ego-networks in coauth-DBLP on star ego-networks in email-Avocado and threads-ask-ubuntu. Our model performed better than random guessing, achieving a prediction accuracy of 83\% on email-Avocado and 72\% on threads-ask-ubuntu. This implies that our model has found principles that apply generally across different domains.

\begin{table}[]
\centering
\caption{\textbf{Classifier Results}}
\begin{tabular}{|c|c|c|l|} \hline
 & Star & Radial & Contracted\\ \hline
coauth-DBLP & $0.93 \pm 0.01$ & $0.91 \pm 0.01$ & $0.85 \pm 0.01$\\ \hline
email-Avocado & $0.84 \pm 0.09$ & N/A & N/A \\ \hline
threads-ask-ubuntu & $0.72 \pm 0.05$ & N/A & N/A \\
\hline\end{tabular}
\label{table:ml}
\end{table}

\subsection{Reconstruction Algorithm}
\noindent We now get to our temporal reconstruction algorithm. In brief, we define a hill climbing algorithm that iteratively swaps all pairs of simplices, applies our supervised method to each resultant ordering, and selects one of the orderings that is more likely to be correctly sorted than the current ordering. This is repeated until no pairwise swaps improve likelihood, in which case this process repeats itself on another randomized ordering. Once we have a selection of improved orderings, the algorithm selects the ordering with highest likelihood. \par 
The accuracy of our model is measured by how many pairs of simplices in the predicted order are out of order with respect to the true answer. That is, of all $m \choose 2$ pairs of simplices $(e_i, e_j)$, what fraction of these pairs appear in the correct time order in the predicted ordering? If we were to guess at random, we would expect to find half of all pairs to be in order; so if a predicted ordering had significantly more than half of its $m \choose 2$ pairs in order, this suggests that the prediction algorithm is capturing significant patterns regarding the temporal structure of hypergraph ego-networks. \par

We provide the pseudocode of our temporal reconstruction algorithm that is described in greater detail in Algorithm 1. The algorithm only requires three parameters: \textit{ego-network $\pi_0$}, \textit{supervised model $M$}, and \textit{number of total iterations $T$}. It then repeats the following steps $T$ times, using a counter variable $i$ initialized to 1 in order to iterate the algorithm until $T$.
\begin{enumerate}
    \item Set $\pi_0(i)$ to a randomly shuffled version of $\pi_0$.
    \item Set $\pi := \pi_0(i)$.
    \item Swap every pair of simplices in $\pi$ and apply $M$ to each resultant ego-network.
    \item For each resultant ordering, if the probability that the resultant ordering is correctly sorted is higher than that of the previous ordering, we will save this ordering.
    \item For all resultant orderings with improved likelihood, select one at random, and set it to $\pi$.
    \item Repeat steps 2-5 until no pair of swapped simplices will improve likelihood, in which case we save the current ordering, increase $i$ by one, and go to step 1.
\end{enumerate}

\begin{table}[t]
\centering
\caption{\textbf{Reconstruction Algorithm Results. We sample 100 ego-networks for each dataset and ego-network type, and report the out of order accuracy for each. We set T = 10.}}
\begin{tabular}{|c|c|c|l|} \hline
 & Star & Radial & Contracted\\ \hline
coauth-DBLP & $0.65 \pm 0.08$ & $0.56 \pm 0.05$ & $0.65 \pm 0.08$\\ \hline
email-Avocado & $0.63 \pm 0.11$ & N/A & N/A \\ \hline
threads-ask-ubuntu & $0.70 \pm 0.07$ & N/A & N/A \\
\hline\end{tabular}
\label{table:reconstruction}
\end{table}

This process is repeated $T$ times until we have a set of improved orderings, at which point we select the ordering with the highest probability of being correctly ordered. \par
Table \ref{table:reconstruction} shows the results of our algorithm on each dataset on a sample of 100 simplices of length between 10 and 15 for email-Avocado, of length 10 for threads-ask-ubuntu, and 20 for coauth-DBLP. As stated previously, a random guessing baseline would expect to find half of all pairs to be out of order, and as a result would achieve an accuracy of 50\%. Another naive but intuitive baseline that uses a more principled heuristic than random ordering would be to sort all ego-networks by increasing simplex size. This baseline achieves results of approximately 50\% across all datasets and ego-network types. Our algorithm's results show a non-trivial improvement over both baselines. We attribute this to the set of powerful features used by our learning method, which take advantage of key ideas such as alter-networks, the intersection size between adjacent simplices, and the time at which user nodes typically arrive in radial and contracted ego-networks.

\subsection{Theoretical Bounds}

It is interesting to ask whether any theoretical guarantees can be made for the quality of hypergraph orderings found by this type of local search.
In general, it is difficult to say anything formal about local search using the trained model, given that we do not have a succinct description of the model.
However, given that maximizing average intersection size (among consecutive simplices in order) serves as an effective heuristic for the ordering problem, we can achieve some insight into the power of local search by proving an analogous result for local search to maximize average intersection size.

We'll quantify the performance by two 
parameters of the instance: $c$, equal to the maximum size of any simplex
in the instance; and $d$, equal to the maximum number of 
simplices that
any one node occurs in.
We also preprocess the instance by deleting all elements that
occur in at most one simplex, since none of these elements can 
contribute to the average intersection size.
We can then delete any simplices that become empty as a result.
In what follows, we will therefore assume without loss of generality
that all simplices are non-empty, and
each element occurs in at least two simplices.

We consider an arbitrary local search algorithm that swaps pairs of simplices as long as the swap strictly increases the objective function of average intersection size.
A local optimum is an ordering at the end of this process, when no swap strictly increases the objective.
We now prove an approximation guarantee for local search.

\begin{theorem}
For any local optimum, the average intersection size is at least $1/(2c^2 d)$ times the average intersection of the globally optimal solution.
\end{theorem}

\proof{
Since each simplex has size $\leq c$,
the optimal solution (with order $\pi$) has average intersection size at most
$$\frac{1}{m-1} \sum_{i=1}^{m-1} |s_{\pi(i)} \cap s_{\pi(i+1)}| \leq \frac{1}{m-1} \sum_{i=1}^{m-1} c = c.$$
Now consider a locally optimal solution; assume
for notational simplicity that it orders the simplices
as $s_1, s_2, \ldots, s_m$.
We partition the indices $1 \leq i \leq m-1$ into two sets:
the set $A$ consisting of indices $i$ such that 
$|s_{i-1} \cap s_i|$ and $|s_i \cap s_{i+1}|$ are both 0,
and 
the set $B$ consisting of all other indices $i$.

Next, let $B'$ be the set of all indices $i$ such that
$|s_i \cap s_{i+1}| > 0$.
We have $B' \subseteq B$;
also, if $i \in B$, then at least one of $i-1$ or $i$ is in $B'$,
from which it follows that $|B| \leq 2 |B'|$.
Since each index $i \in B'$ contributes
at least $1$ to the total intersection size, the average
intersection size
in our locally optimal solution is at least $\frac{1}{m-1} |B'| \geq \frac{1}{2(m-1)} |B|$.

Now we come to the key step.
For each $i \in A$, consider an arbitrary $u_j \in s_{i+1}$.
The simplex $s_i$ does not contain $u_j$ (since otherwise
we would have $|s_i \cap s_{i+1}| > 0$, contradicting $i \in A$).
But $u_j$ occurs in at least two simplices; let $s_h$ be
another simplex in which it occurs.
We cannot have $h \in A$, since then swapping $s_h$ and $s_i$ would strictly increase the average intersection size, contradicting local optimality.
Thus $h \in B$.

We now charge index $i \in A$ to index $h \in B$.
$s_h$ has $\leq c$ elements, and each can be charged $\leq d-1$ times, so $h$ can be charged $\leq c(d-1)$ times; hence $|A| \leq c(d-1) |B|$.
We also have $|A| + |B| = m-1$, so 
$|B| \geq \frac{m-1}{1 + c(d-1)} \geq \frac{m-1}{cd}.$

Thus
the locally maximum solution has average intersection size 
$\geq \frac{|B|}{2(m-1)} \geq \frac{1}{2cd}$ while the optimum solution
has total intersection size at most $c$.
The optimum solution therefore has average intersection size
at most $2c^2 d$ times that of the locally optimal solution, completing the proof.
}

\section{Related Work}

\begin{algorithm}[t]
    \SetAlgoLined
    \SetKwRepeat{Do}{do}{while}
    \textbf{Input:} ego-network $\pi_0$, supervised model $M$, and number of total iterations $T$\\
    \textbf{Output:} a predicted ordering $\max_{j=1}^{T} L(j)$  \\
    $L \leftarrow \emptyset$ \\
    $i \leftarrow 1$ \\
    \While{$i \leq T$}{
      $\pi_0(i) \leftarrow random(\pi_0)$ \\
      $\pi \leftarrow \pi_0(i)$ \\
      \Do{$U \neq \emptyset$}{ $U \leftarrow$ all orders $\pi'$ obtained by one swap from $\pi$ where $M(\pi') > M(\pi)$ \\
      $\pi \leftarrow$ random choice from $U$ \\
      }
      $L(i) \leftarrow \pi$ \\
      $i \leftarrow i + 1$
    }
    \caption{Temporal Reconstruction hill climbing algorithm for sorting unordered hypergraph ego-networks}
\end{algorithm}

\noindent A large volume of past work has been done on the evolution of global dyadic graphs \cite{leskovec2007graph, leskovec2005graphs, leskovec2010kronecker}. The evolution of dynamic systems that model higher-order interactions using hypergraphs has also been previously investigated \cite{benson2018simplicial, kook2020evolution, benson2016higher, do2020structural}. \cite{benson2018simplicial} studies the temporal evolution of global hypergraph datasets in the context of simplicial closures and link prediction, and also looks at predicting system domain using higher-order ego-networks. In contrast, our paper attempts to understand the temporal evolution of local hypergraph ego-networks, extracting higher-order ego-network features to predict evolution rather than system domain. \par
\cite{benson2018simplicial} also mentions the projected graph, the encoding of higher-order information as a traditional dyadic graph. Significant information loss occurs when hypergraphs that model higher-order interactions are converted into the projected graph. Thus, we do not use the projected graph in our paper to analyze higher-order networks. \cite{kook2020evolution} examines temporal properties of global hypergraphs in order to realistically generate hypergraphs. In contrast, our paper instead focuses on modelling local hypergraph structure.\par
Dyadic ego-networks have been used to model local interactions across various fields, including social and co-authorship ego-networks \cite{arnaboldi2017online}. Temporal dyadic ego-network evolution has also been frequently studied \cite{aiello2017evolution, arnaboldi2016analysis}. There is also past work on using machine learning methods to understand the structural patterns in pairwise ego-networks \cite{mcauley2012learning}. In \cite{aiello2017evolution}, dyadic ego-networks from social media datasets are analyzed in order to describe common patterns found regarding their growth.
\cite{aiello2017evolution} also finds that dyadic ego-networks tend to rapidly expand at the beginning of their lifetimes, adding many nodes in a short period of time. In our paper, we show the opposite to be true in the co-authorship case, where user nodes typically strengthen their ties with early alters rather than finding new nodes.\par
Benson et al analyze repeat behavior in sequences of sets \cite{benson2018sequences} via a formalism they term the Correlated Repeated Unions (CRU) model.  In our context, we could model a hypergraph ego-network as a sequence of sets, with each set being a simplex.  Both our paper and \cite{benson2018sequences} identify a recency bias, but we consider different questions, with our paper focusing on orderings rather than set composition.
\section{Discussion}
\noindent In this paper, we have proposed the study of hypergraph ego-networks, a structure that can be used to model higher-order interactions involving a single node and its alters. We define three ego-network types: star, radial, and contracted ego-networks, each modeling different higher-order interactions surrounding the user node. We have examined higher-order interactions across three domains: co-authorship (via a collection of publications \textit{coauth-DBLP}), communication (via a collection of emails \textit{email-Avocado}), and online threads (via a collection of users participating in a thread \textit{threads-ask-ubuntu}). The coauth-DBLP and threads-ask-ubuntu datasets can be found at: \url{https://www.cs.cornell.edu/~arb/data/}, and email-Avocado can be found at \url{https://catalog.ldc.upenn.edu/LDC2015T03}. Our code can be found at \url{https://github.com/Cazamere/hypergraph-assembly}. \par
Our work introduces several key observations that subsequently inform a set of prediction algorithms to accurately reconstruct hypergraph ego-networks. The most powerful of these is the \emph{alter-network spread}, the average temporal distance between adjacent simplices in each alter-network of an ego-network. We find that alter-networks possess strong temporal locality, as they tend to occupy very defined sets of proximate timestamps in the ego-network they belong to.  This notion of temporal locality within a collection of sets suggests interesting connections to other contexts where similar locality phenomena arise.  Perhaps most directly, it would be interesting to consider connections to the well-known principle of locality of reference in computer systems, which is also based on the idea that occurrences of particular elements in long access patterns are clustered in time \cite{tanenbaum2015modern}.  And as a slightly more distant but intriguing connection, notions of locality in set systems form the underpinning for fundamental combinatorial questions about high-dimensional polyhedra, where the ordering of vertices by breadth-first search obeys a form of locality with respect to the facets that contain them \cite{eisenbrand2009diameter}. \par
Interestingly, we also find that radial ego-networks in coauth-DBLP contain a large and constant fraction of user simplices, regardless of size, as user nodes arrived into their radial ego-networks around or slightly before the fifth timestamp. However, this is not the case for contracted ego-networks, where an approximately linear relationship is observed between the time a user node arrives into their ego-network and the size of the ego-network. \par
Finally, we also propose the temporal reconstruction of hypergraph ego-networks as a benchmark problem for models that aim to predict the local temporal structure of hypergraphs. As a solution, we propose a supervised deep learning method to learn if a given hypergraph ego-network is correctly ordered in time. Next, we define a hill climbing algorithm that is given our learning method and a shuffled ego-network. Our model significantly outperforms several baselines by incorporating various structural patterns found in hypergraph ego-networks across multiple domains such as intersection density and alter-network spread. We envision that our model may act as a foundation for future work that aims to further understand the local, temporal structure of higher-order interactions. \par

\nocite{lee2021hyperedges,lee2020hypergraph}

\bibliographystyle{abbrv}
\bibliography{refs-short-v02}

\end{document}